\documentclass[%
 reprint,
 amsmath,amssymb,
 aps,
 showkeys,
]{revtex4-2}

\usepackage{graphicx}
\usepackage{dcolumn}
\usepackage{bm}
\usepackage{xcolor}
\usepackage[T1]{fontenc}
\usepackage[normalem]{ulem}
\usepackage{amssymb}

\newcommand{\eself}{\varepsilon_1}
\newcommand{\eseries}{\varepsilon_2}
\newcommand{\de}{\Delta\varepsilon}
\newcommand{\rhop}{\rho_\mathrm{p}}

\begin{document}

\preprint{APS/123-QED}

\title{Stochastic size control of self-assembled filaments}

\author{Maximilian C. H\"ubl}
\email{maximilian.huebl@ist.ac.at}
\author{Carl P. Goodrich}%
\email{carl.goodrich@ist.ac.at}
\affiliation{%
 Institute of Science and Technology Austria (ISTA), Am Campus 1, 3400 Klosterneuburg, Austria
}%

\begin{abstract}
Controlling the size and shape of assembled structures is a fundamental challenge in self-assembly, and is highly relevant in material design and biology.
Here, we show that specific, but promiscuous, short-range binding interactions make it possible to economically assemble linear filaments of user-defined length.
Our approach leads to independent control over the mean and width of the filament size distribution and allows us to smoothly explore design trade-offs between assembly quality (spread in size) and cost (number of particle species).
We employ a simple hierarchical assembly protocol to minimize assembly times, and show that multiple stages of hierarchy make it possible to extend our approach to the assembly of higher-dimensional structures.
Our work provides a simple and experimentally straightforward solution to size control that is immediately applicable to a broad range of systems, from DNA origami assemblies to supramolecular polymers and beyond.
\end{abstract}

\maketitle

One of the hardest qualities to control in the design of self-assembling structures is also among the simplest: size. 
Nevertheless, achieving precise control over a structure's size -- the number of subunits in the structure -- is extremely important in many areas, from biology~\cite{Alberts2015, Rhind.2021, Ginzberg.2015, Hagan.2021} and biomedical applications~\cite{Kim.2005, Rad-Malekshahi.2016, Sigl.2021} to photonic materials~\cite{Vukusic.2003, Dufresne.2009, Satyabola.2025, Hayakawa.2024, Hensley.2023} and nanofabrication~\cite{Aldaye.2008, Pearce.2021, Michelson.2024, Michelson.2025}, because size is often directly tied to the function of the assembled object.
To control the function, one must first control the size.

The challenge is relating the global property of structure size to the local properties of the assembling particles: how should the particles ``know'' when to stop aggregating, and how can that knowledge be communicated via the particles' interactions?
In some approaches to this problem, information is encoded in the particle geometry, which can lead to structures that self-close on controllable length scales.
These strategies have been successfully demonstrated in the assembly of sheets, tubules, shells, or even more complex manifolds~\cite{Sigl.2021, Hayakawa.2022pqs, Videbaek.2024, Duque.2024, Saha.2025}.
Other approaches construct interactions so that the energy of a structure contains terms that scale differently with size, leading to a preferred size that is determined by the competition of these terms.
Such cooperative effects can be mediated through long-range repulsion~\cite{Sciortino.2004, Nguyen.2015}, geometric frustration~\cite{Grason.2016,lenz2017geometrical-1f1,roy2025collective-bd3}, or even frustration between crystalline domains~\cite{koehler2025topological-bb2}.

\begin{figure}
    \centering
    \includegraphics[width=\linewidth]{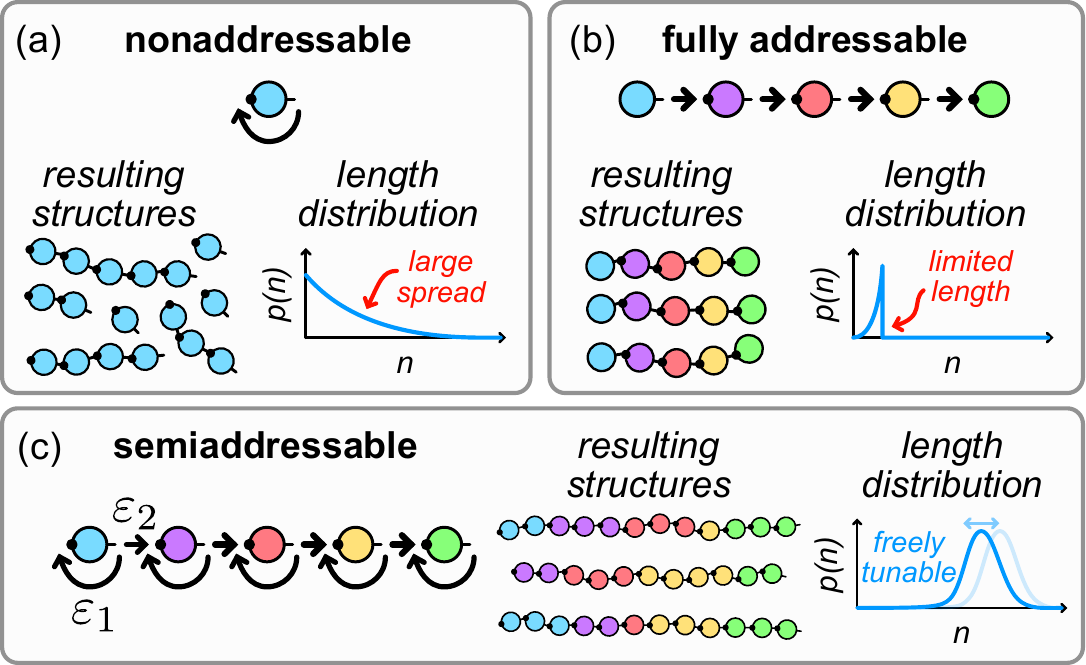}
    \caption{Different filament assembly strategies. (a) Nonaddressable assembly with short-ranged interactions leads to a very broad length distribution. (b) Fully addressable assembly makes it possible to target a specific length, but that length is limited by the number of particle species, making this strategy not scalable. (c) Semiaddressable assembly, where particles bind promiscuously, makes it possible to engineer a narrow and tunable distribution of structure sizes.}
    \label{fig:fig1}
\end{figure}

However, such approaches are particularly challenging for one-dimensional structures such as linear filaments as they allow only limited cooperativity between particles, constraining the potential mechanisms through which information can be transmitted from one end of the filament to the other.
Le Roy et al.~\cite{roy2025collective-bd3} demonstrate a powerful, yet experimentally demanding, approach to 1d size control through geometric frustration with deformable particles, while other approaches exploit complex, non-equilibrium mechanisms that play an important role in biology~\cite{Johann.2012, Melbinger.2012, Striebel.2022, Kuan.2013, Datta.2018}.
Another potential approach is through multiple particle species with specific interactions. Whereas generic single-species assembly leads to a wide distribution of filament lengths (Fig.~\ref{fig:fig1}a), Fig.~\ref{fig:fig1}b shows a scheme for designing particles to assemble in a precisely ordered sequence. However, while this can lead to size control~\cite{Murugan.2015}, it requires as many particle species as the desired filament is long, which makes experimental realizations highly costly and impractical, even for only moderately long filaments.

Here, we show how to achieve the size-control benefits of such addressable assembly but with a fixed and reasonable number of species by exploiting configurational entropy. 
Using multiple particle species with programmable interactions, we assemble filaments whose length can be freely tuned by adjusting particle concentrations or binding energies -- independently of the number of particle species.
This tunability allows us to smoothly explore design trade-offs between assembly quality (spread in size) and cost (number of particle species), where nonaddressable assembly (low cost, low quality) and fully-addressable assembly (high cost, high quality) are recovered as limiting cases.
We find that high-quality assembly of long filaments requires long assembly times, and we propose a simple hierarchical assembly protocol that can speed up the assembly process by over five orders of magnitude, thereby making high-quality assembly feasible on experimentally accessible timescales.
Since our approach does not rely on any secondary interactions beyond specific bond formation, it is directly applicable to many programmable self-assembly platforms, such as DNA-based systems~\cite{Hayakawa.2024, Videbaek.2024, Saha.2025, He.2020, Mirkin.1996, Rogers.2015, Wang.2012, Wang.2015, Valignat.2005, Jacobs.2025}, systems with shape-complementary interactions~\cite{Sacanna.2010, Sacanna.2013}, and certain \emph{de novo} proteins~\cite{Huang.2016, King.2012} or other supramolecular polymers~\cite{Greef.2008, Aida.2012, Eikelder.2019}.

The fundamental quantity of interest in this paper is the equilibrium length distribution $p(n)$ of the filaments, and it is our goal to design the mean $\langle n \rangle$ and width $\sigma$ of this distribution.
The challenge of size control becomes apparent if we look at nonaddressable (single species) filaments, as shown in Fig.~\ref{fig:fig1}(a).
The equilibrium length distribution can be computed from the structure partition functions~\cite{Murugan.2015, Klein.2018, curatolo.2023, Holmes-Cerfon.2013, Holmes-Cerfon.2016, Huebl.2025, Huebl.2025b} and is given by
\begin{equation}\label{eq:nonaddressable}
    p_\mathrm{na}(n) = \frac{e^{-\lambda n}}{C_\mathrm{na}} \,,
\end{equation}
where $\lambda = - \beta(\varepsilon + \mu)$ depends on the binding energy $\varepsilon$, chemical potential $\mu$ and inverse temperature $\beta = 1/kT$, and where the normalization constant $C_\mathrm{na}$ is proportional to the partition function of the system (see Supplementary Information).

The important point is that the equilibrium assembly of a single particle species is characterized by a single control parameter $\lambda > 0$.
Adjusting $\lambda$ allows one to tune the characteristic length of the aggregates, but since the length distribution is exponential~\footnote{Strictly speaking, the distribution is \emph{geometric}, since $n$ is an integer and $1 \leq n < \infty$.}, the width of the distribution is always comparable to the mean, $\sigma \sim \langle n \rangle$.
Achieving size control, {\it i.e.} lowering $\sigma$ below this baseline, is not possible without some additional design strategy~\cite{Hagan.2021, brenner2017}.
As discussed above, one possibility is to introduce multiple particle species with specific interactions, which allows one to precisely select the length of the filaments~\cite{brenner2017, Murugan.2015} (Fig.~\ref{fig:fig1}(b)).
However, a direct, fully addressable design requires the number of species and interactions to grow linearly with the desired target length, making this approach costly and impractical in experiments~\footnote{This remains true if the two-fold symmetry of a filament is exploited, which cuts down the required particle species by half, but leaves the prohibitive linear scaling unaffected.}.

To avoid both of these limitations, we now introduce addressability in a more careful way.
The resulting ``semiaddressable'' design~\cite{Huebl.2025} will allow us to freely tune the mean and width of the length distribution.
Our design consists of $m$ particle species that each can bind to themselves to form single-species filaments, just as the nonaddressable system discussed above.
However, there are additional interactions that allow the ``right side'' of particle species $i$ to bind to the ``left side'' of species $i+ 1$, so that the single-species filaments can be joined sequentially, as shown in Fig.~\ref{fig:fig1}(c).

The idea behind this design becomes clear if we imagine for the moment that the assembly proceeds hierarchically, so that the $m$ species initially just bind to themselves and exclusively form single-species filaments.
Once the single-species filaments have reached equilibrium, we freeze them and turn on the cross-species interactions, letting the single-species filaments combine in groups of (up to) $m$ to form multi-species filaments.
The length of a multi-species filament is the sum of $m$ single-species filament lengths, which are uncorrelated and exponentially distributed random variables, as discussed above.
Appealing to the central limit theorem, we may therefore expect that the length distribution of the multi-species filaments approaches a normal distribution as the number of species $m$ is increased --- if this is the case, the mean and variance of the filament length can be precisely and independently controlled.

We now make these statements rigorous and show that this design works as intended even if the assembly does not proceed hierarchically.
We make the simplifying assumption that all $m$ particle species are supplied at the same chemical potential $\mu$ and we denote the same-species binding energy by $\eself$ and the cross-species binding energies by $\eseries$.
We can immediately write down the length distribution by noting that the probability of observing a length-$n$ filament is proportional to the number of ways it can be decomposed into $k$ single-species filaments.
Summing over all the ways $k$ sequential species can be picked out of $m$ species, and keeping track of the binding energies and chemical potentials, leads us to the semiaddressable length distribution:
\begin{equation}\label{eq:dist_full}
    p_\mathrm{sa}(n) = \frac{e^{-\lambda n}}{C_\mathrm{sa}} \sum_{k=1}^{m}(m-k+1) e^{k\beta\de} \binom{n-1}{k-1} \,,
\end{equation}
where $\de = \eseries - \eself$ and $\lambda = -\beta(\eself + \mu)$.

Analytic expressions for the normalization constant $C_\mathrm{sa}$, mean $\langle n \rangle$, and standard deviation $\sigma$ of $p_\mathrm{sa}(n)$ can be found in Supplementary Information, but they are long and not very illuminating.
However, we \textit{can} gain valuable intuition by considering the limit where the mean $\langle n \rangle$ of the distribution is much larger than the number of species $m$.
In this limit, we can neglect partially assembled filaments (i.e., filaments that do not contain all $m$ species) and the general distribution can be simplified considerably.
Specifically, if 
\begin{equation}\label{eq:semi_limit}
    \langle n \rangle \gg 2me^{-\beta\de} \,,
\end{equation}
the assembly is dominated by ``complete'' filaments, and we can neglect all but the $k=m$ term in the sum of Eq.~\eqref{eq:dist_full}.
To simplify further, we also approximate the binomial coefficient as $\binom{n-1}{k-1} \approx (n - k/2)^{k-1} / (k-1)!$, after which the length distribution assumes the simpler form
\begin{equation}\label{eq:dist_limit}
    p_\mathrm{sa}(n) \approx \begin{cases}
                              0 \quad \text{if $n < \frac{m}{2}$} \,, \\
                              \frac{\lambda^m}{(m-1)!} \left(n - \frac{m}{2}\right)^{m-1} e^{-\lambda \left(n - \frac{m}{2}\right)}  \quad \text{else}\,.
            \end{cases}
\end{equation}

This limiting distribution is an \emph{Erlang} distribution, shifted by $m/2$.
The \emph{Erlang} distribution describes sums of exponentially distributed variables, so it is not surprising that our general distribution reduces to it in the limit where all filaments are complete.
Note that the limiting distribution is independent of the cross-species binding energy $\eseries$ because we are in a regime where filaments always contain $m-1$ cross-species bonds.
The mean and standard deviation of the limiting distribution have the convenient analytical expressions
\begin{equation}\label{eq:moments}
    \langle n \rangle = m \left[\frac{1}{2} + \frac{1}{\lambda}\right] \,, \quad \sigma = \frac{\sqrt{m}}{\lambda} \,.
\end{equation}

\begin{figure}
    \centering
    \includegraphics[width=\linewidth]{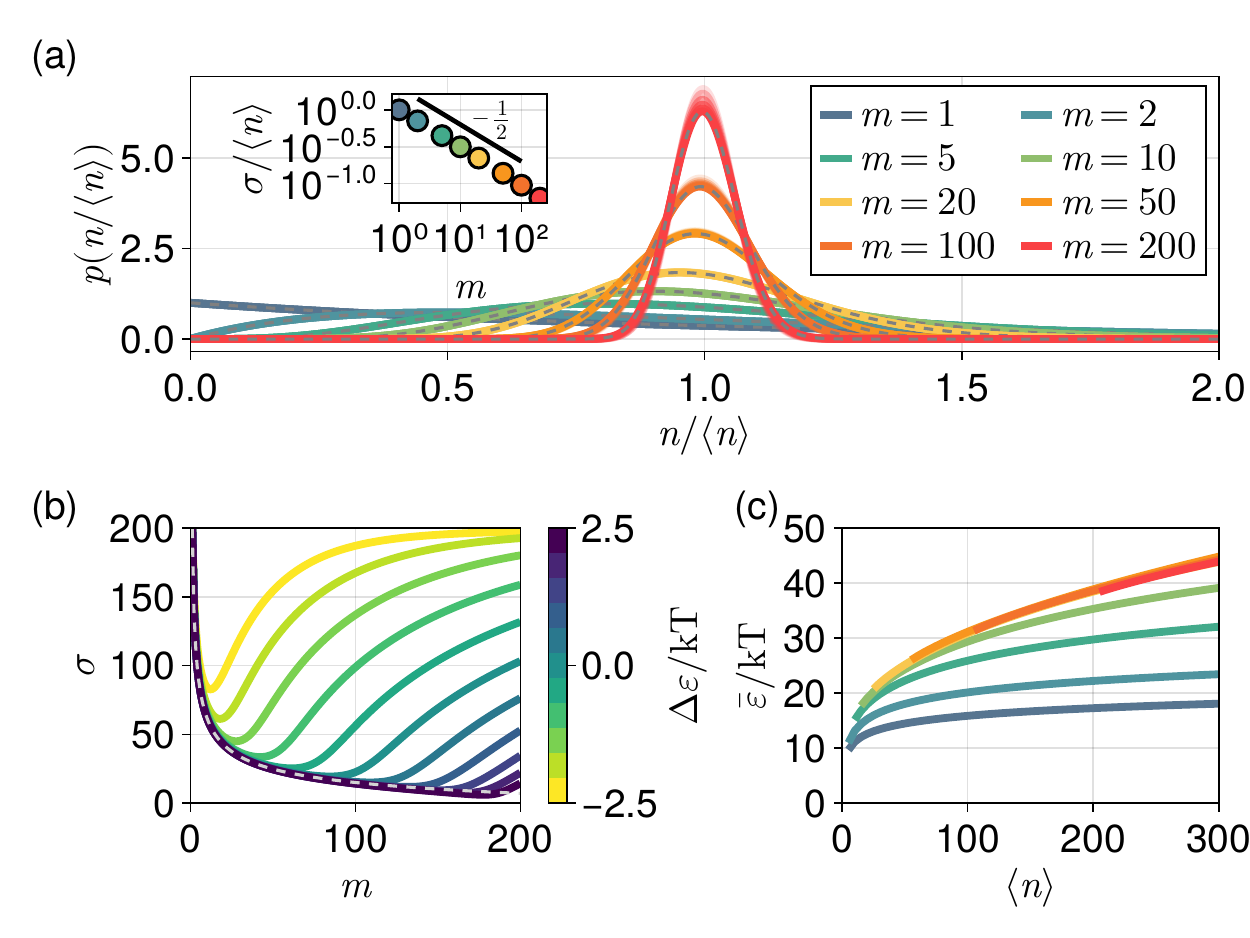}
    \caption{Equilibrium properties of semiaddressable filaments. (a) Scaled equilibrium length distributions for different numbers of particle species $m$. Target lengths range from $\langle n \rangle = 500-1000$, and are well approximated by the large--$\langle n \rangle$ limit.
    The distributions corresponding to different target lengths are shown in different lightness and converge to the limiting distribution given by Eq.~\eqref{eq:dist_limit} (dashed gray lines).
    Inset shows how the relative peak width scales with $m$ (colored points); the black line shows theoretical $m^{-1/2}$ scaling.
    The volume fraction is $\phi=0.1$ and $\Delta\varepsilon=0$.
    (b) Peak width as a function of the number of species $m$, for different $\Delta\varepsilon$ at fixed $\langle n \rangle = 200$. The dashed gray line shows the limiting expression for $\sigma$ given in Eq.~\eqref{eq:moments}.
    (c) The average binding energy $\bar{\varepsilon} = (\eself + \eseries) / 2$ required to reach a given $\langle n \rangle$ while maintaining a peak width of $\sigma = \langle n \rangle / \sqrt{m}$, for different numbers of species. Colors are the same as in (a).}
    \label{fig:fig2}
\end{figure}

Figure~\ref{fig:fig2}(a) shows that the semiaddressable length distribution has a pronounced peak whose width decreases as the number of species is increased, which confirms the effectiveness of our design.
Moreover, in the large--$\langle n \rangle$ limit, the distributions for different target lengths can be collapsed onto a single curve; Figure~\ref{fig:fig2}(a) shows the scaled length distribution for different numbers of particle species.
We view the ratio of standard deviation to mean, $\sigma / \langle n \rangle$, which we call the ``relative peak width'', as a measure of the quality of the assembly -- the smaller the relative peak width, the higher the quality.
The inset in Fig.~\ref{fig:fig2}(a) shows that, in the large--$\langle n \rangle$ limit, the relative peak width scales as $m^{-1/2}$, which is expected from the central limit theorem and also follows from Eq.~\eqref{eq:moments} if $\langle n \rangle \gg m$.
Figure~\ref{fig:fig2}(a) makes it clear that our approach allows us to freely tune the average filament length independently of the number of species, and that adding more species to the system increases the quality of the assembly.
To put this in perspective: in DNA origami based systems, for example, it is possible to create over 20 distinct particle species~\cite{Videbaek.2024, Hayakawa.2022pqs}, which would lead to a size dispersity of roughly $20\%$.

Equation~\eqref{eq:semi_limit} implies that the value of $\de$ significantly affects whether the large--$\langle n \rangle$ limit can be reached.
The consequences of this are shown in Fig.~\ref{fig:fig2}(b), where the peak width is shown as a function of the number of species, for different values of $\de$ and at constant $\langle n \rangle = 200$.
For small $\de$, the assembly quality deteriorates if too many species are used.
$\de$ controls how favorable it is for different single-species filaments to join together, so if $\de$ is too low, not all single-species filaments can aggregate, and the assembly is dominated by incomplete filaments that contain less than $m$ segments, which negatively impacts assembly quality.

Considering the energies required for successful assembly points us to important practical limitations.
Figure~\ref{fig:fig2}(c) shows the binding energies necessary to reach a given target length while maintaining the relative peak width from the large--$\langle n \rangle$ limit, $\sigma / \langle n \rangle = m^{-1/2}$.
It shows that assembling long filaments with high quality can easily require binding energies greater than $30\,\mathrm{kT}$.
Since the binding energies determine the timescale on which bonds can break and the system can equilibrate, 
this suggests that there is an unavoidable trade-off between filament length, quality, and assembly time.

\begin{figure}
    \centering
    \includegraphics[width=\linewidth]{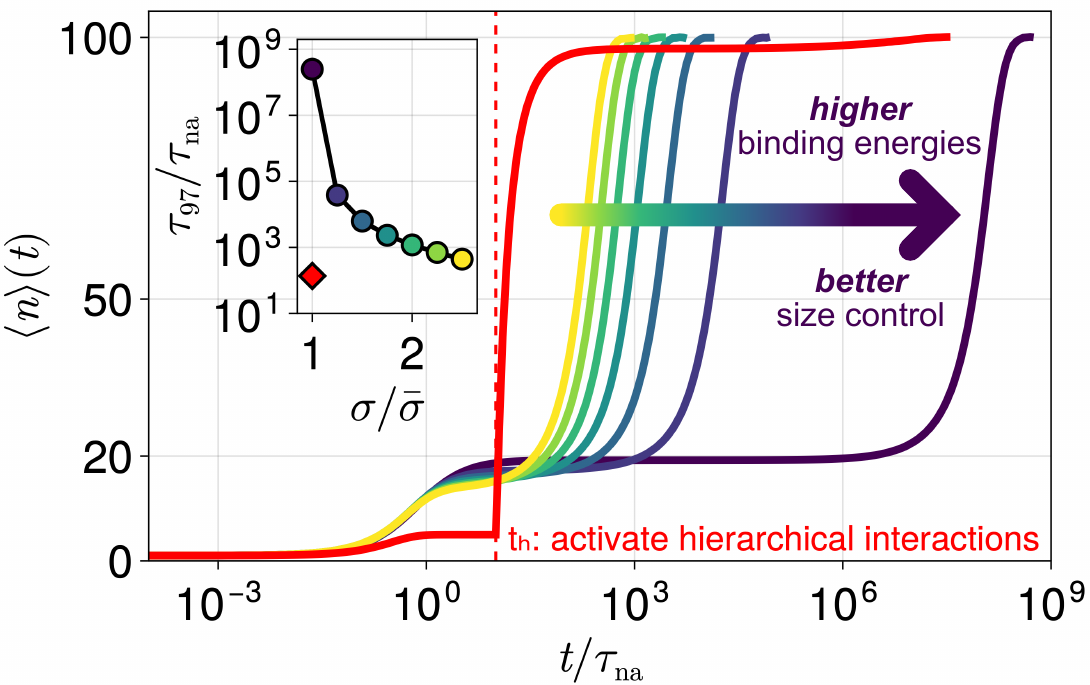}
    \caption{Kinetics of semiaddressable filaments. Time dependent average ensemble length for $m=10$ species and a target length of $\langle n \rangle(t = \infty) = 100$, for target peak widths $\sigma / \bar\sigma$ ranging from $1$ (dark purple) to $2.5$ (yellow), which are achieved by varying the binding energies. We use $\bar\sigma = \langle n \rangle / \sqrt{m}$, which is approximately equal to the peak width of the limiting distribution, Eq.~\eqref{eq:moments}, as a reference. The red curve shows hierarchical assembly, targeting $\sigma / \bar\sigma = 1$. Time is measured in units of the nonaddressable equilibration time $\tau_\mathrm{na}$.
    (\textit{inset}) Assembly time $\tau_{97}$ as a function of the target peak width for direct assembly (circles) and hierarchical assembly (diamond), showing the trade-off between assembly quality and time. For non-hierarchical systems, colors indicate peak width and are identical to the ones used in the main panel.}
    \label{fig:fig3}
\end{figure}

To understand this trade-off better, we investigate the equilibration process by performing Markov-based kinetic simulations, as described in \textit{Appendix}.
The results of these simulations are shown in Fig.~\ref{fig:fig3}, which shows the time-dependent average ensemble length $\langle n \rangle(t)$ for $m=10$ particle species, and a target length of $\langle n \rangle(t = \infty) = 100$.
Different curves are obtained at varying binding energies, which were chosen to reproduce specific peak widths $\sigma$ (this was done by numerically inverting the analytical relations shown in Supplementary Information).
We use the equilibration time of the nonaddressable system, $\tau_\mathrm{na}$, as a reference timescale.
Figure~\ref{fig:fig3} shows that the assembly time of an addressable multi-species system can be orders of magnitude higher compared to a nonaddressable system, and it confirms that the target peak width $\sigma$ has a strong impact, since smaller $\sigma$ requires higher binding energies.

Given that nonaddressable equilibration times are often on the order of minutes to hours, the long equilibration times required for high-quality, multi-species assembly shown in Fig.~\ref{fig:fig3}(a) may be prohibitively long.
To promote the assembly of high-quality distributions on experimentally feasible timescales, we now consider a hierarchical strategy like the one we used to motivate our design in the first place: we first form the single-species filaments, and only later turn on the cross-species interactions that allow filaments to join sequentially.

In this hierarchical approach, we initially choose $\eself$ such that all $m$ species assemble single-species filaments of average length $\langle n_\mathrm{na}\rangle = \langle n \rangle / (2m)$~\footnote{The length of $\langle n \rangle / (2m)$ is necessary, since there is a roughly 50\% chance a single-species segment binds with another one of the same species before encountering a filament of the subsequent species.}, while disallowing any cross-species bonds from forming. Here, $\langle n \rangle $ is the desired eventual average length of the fully assembled filaments. 
After a time $t_\mathrm{h} \geq \tau_\mathrm{na}$, we turn on the cross-species interactions, and set both $\eself$ and $\eseries$ to the values required for equilibrium assembly with desired $\langle n \rangle$ and $\sigma$.
Experimentally, such an instantaneous change in interactions could, for example, be achieved through a change of temperature~\cite{Hayes.2021} or through the addition of DNA linkers~\cite{Xiong.2009, Lowensohn.2019}.
In Supplementary Information, we also consider a stronger form of hierarchy where we freeze the single-species filaments after the initial equilibration period, which yields similar results.

The hierarchical assembly kinetics are shown as the red curve in Fig.~\ref{fig:fig3}, demonstrating that hierarchical assembly leads to a drastic speed-up compared to all-at-once assembly.
The inset to Fig~\ref{fig:fig3} shows the measured assembly time, $\tau_{97}$, defined as the first time at which $\langle n \rangle(\tau_{97}) = 0.97\langle n \rangle$, as a function of the target peak width, showing a stronger-than-exponential dependence for non-hierarchical assembly.
Compared to this, hierarchical assembly times are more than five orders of magnitude lower.

Note though that the hierarchical system is not completely equilibrated at $\tau_{97}$.
Once the hierarchical step at $t_\mathrm{h}$ is completed, the hierarchical system assembles under the same conditions as the highest quality non-hierarchical system (dark purple curve in Fig.~\ref{fig:fig3}) and the final relaxation to $\langle n \rangle = 100$ is thus governed by a similar, long timescale.
However, hierarchical assembly gets us very close to the equilibrium state very quickly -- for most practical purposes, the hierarchical length distribution at $\tau_{97}$ will be indistinguishable from the equilibrium distribution, Eq.~\eqref{eq:dist_full}.
See Fig.~S1 in Supplementary Information for a comparison.

Our results show that size control of linear filaments can be achieved by carefully designing specific, short-range interactions between an experimentally reasonable number of species, following the semiaddressable scheme presented in Fig.~\ref{fig:fig1}(c). This leads to an equilibrium distribution of filament lengths whose mean and width can be freely tuned. We have identified two fundamental tradeoffs that govern optimal design. The first tradeoff is between quality and cost: as shown by Fig.~\ref{fig:fig2}(a), higher cost (higher number of species $m$) leads to higher quality (lower $\sigma/\left<n\right>$). The second tradeoff is between quality and equilibration time: for fixed cost (fixed $m$), both quality and equilibration times depend on binding energies, with higher quality requiring higher equilibration times (Fig.~\ref{fig:fig3}). This second tradeoff can be mitigated through a simple hierarchical protocol, as discussed. 

Size control is achieved through a competition of bulk free energy and configurational entropy.
This entropy comes from the possible arrangements of the domain walls separating different particle species: longer filaments allow for a larger number of domain wall configurations [which are counted by the binomial coefficient in Eq.~\eqref{eq:dist_full}], and are therefore entropically favored.
On the other hand, adding particles to a filament comes with a free energy penalty quantified by $\lambda$.
The balance between these two effects, which scale differently with structure size, leads to a preferred filament length that is continuously tunable by varying binding energies or particle concentrations.

We have focused our analysis on the assembly of one-dimensional filaments, which pose a particular challenge for many other strategies for size control. However, our approach could be extended to form higher-dimensional structures with controlled size and shape. 
In \emph{Appendix}, we show how pairing our approach with additional layers of hierarchical assembly makes it possible to control the size and shape of two-dimensional sheets.
In short, we introduce ``vertical'' interactions so that size-controlled filaments stack on top of each other, which allows us to iterate our approach and also achieve size control in the second dimension. 
By introducing multiple bond types for these vertical interactions, the height $h$ and width $w$ of the 2d sheet can be controlled in a way that is closely analogous to the 1d case.
We demonstrate this in Fig.~\ref{fig:fig4} in \emph{Appendix}, which shows an example where $\left<w\right> \approx \left<h\right> \approx 100$, and with $\sigma_w/\left<w\right> \approx 0.05$ and $\sigma_h/\left<h\right> \approx 0.38$.
This is only one of many potential ways our approach could be generalized to higher dimensions and we expect that exploiting the trade-off between binding energy and configurational entropy could be a fruitful approach to steering self-assembly outcomes in a variety of contexts.

\begin{acknowledgements} We thank Maitane Muñoz-Basagoiti for helpful discussions.
The research was supported by the Gesellschaft f\"ur Forschungsf\"orderung Nieder\"osterreich under project FTI23-G-011. 
\end{acknowledgements}

\section*{Appendix}
\paragraph*{Filament assembly kinetics ---}
In simulations, we treat the assembly process as a Markov chain, consisting of $N = 5\times10^7$ particles in a volume $V$, and simulate it using the Gillespie algorithm~\cite{Gillespie.2007}.
We assume that two particles (more precisely: two binding sites) encounter each other at a constant rate $\alpha/V$, and that every encounter between compatible species leads to the formation of a bond.
The bonds between particles of the same species break at a constant rate $\delta_1$, and the bonds between particles of different species break at a rate $\delta_2$.
These rate constants are related to the binding energies via $\alpha / \delta_1 = e^{\beta\eself} / (8\pi^2 \, \phi_0)$ and $\alpha / \delta_2 = e^{\beta\eseries} / (8\pi^2 \, \phi_0)$~\cite{Kampen1992}, where $\phi_0$ is an arbitrary reference concentration (see Supplementary Information).
For single-species assembly, this model describes the stochastic kinetics of Flory-Stockmayer polymerization~\cite{Blatz.1945, Stockmayer.1943, Dongen.1984}.
As mentioned in the main text, we use the single-species (nonaddressable) equilibration time $\tau_\mathrm{na} = (\alpha \delta_1 \rhop)^{-1/2}$ as a reference time scale, where $\rhop = N / V$ is the total particle concentration.
To accurately compare the equilibrium calculations with the kinetic simulations, we analytically compute the equilibrium particle concentrations of every species as a function of binding energies and chemical potential, and then initialize the simulation with the corresponding concentrations.
See Supplementary Information for further details.

We can gain a more intuitive understanding of the long equilibration times reported in Fig.~\ref{fig:fig3}, if we consider the assembly kinetics more closely.
At early times, a system consists predominantly of free monomers and bond-breaking events are rare compared to binding events, which means that bonds can be viewed as unbreakable early on.
Depending on particle concentrations, any particle has a roughly $50\%$ chance of encountering either another particle of the same species or one of the subsequent species, which means that the expected filament length at the end of this initial growing phase is about $2m$.
Note that this is not exact since we use nonuniform particle concentrations to maintain constant $\mu$, see Supplementary Information for the exact particle concentrations used.

After this initial growth, filaments become kinetically arrested.
While they can still grow from the ends, to grow one of the segments in the bulk, the filament has to break open first.
Equilibrating the lengths of all filament segments therefore requires a large number of bond breaking events, which means that the approach to equilibrium is governed by the bond breaking timescales $\delta_1$ and $\delta_2$.
These timescales are determined by the binding energies required to assemble filaments with a given length $\langle n \rangle$ and peak width $\sigma$.
Due to the exponential relationship between energies and rates, even a small change in binding energies can have a large effect on the equilibration time.

\begin{figure}
    \centering
    \includegraphics[width=\linewidth]{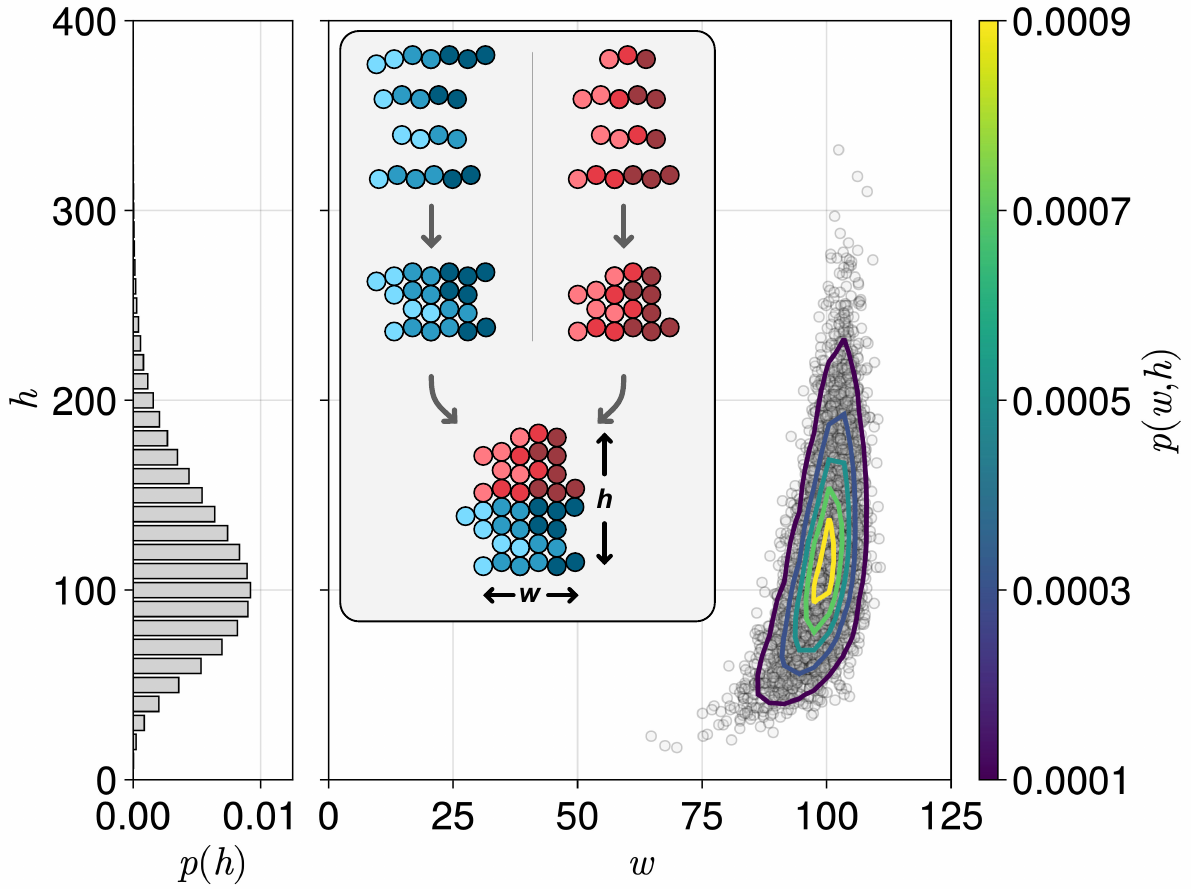}
    \caption{Size distribution of hierarchically assembled two-dimensional sheets. The main panel shows the probability $p(w, h)$ of observing a sheet with width $w$ and height $h$, where width is defined as the mean length of the constituent filaments and height is given by the number of filaments in a sheet. Data points correspond to individual sheets and contour lines show a two-dimensional histogram. The panel on the left shows the marginal distribution of sheet heights $p(h)$. 
    Inset shows a cartoon sketch of sheet assembly, starting from two filament species (shown in blue and red, different shades indicate the different particles species needed for the initial filament assembly), each with the same length distribution but distinct vertical binding sites.
    These filaments first come together to assemble single-species sheets, which then in turn combine into multispecies sheets.}
    \label{fig:fig4}
\end{figure} 

\paragraph*{Assembly of two-dimensional sheets ---}
Figure.~\ref{fig:fig4} shows the size distribution of two-dimensional sheets, which are assembled from vertically ``stacked'' filaments.
The assembly protocol employed here is the following: We first assemble filaments consisting of $m_1=10$ species and targeting a peak width of $\sigma = \langle n \rangle / \sqrt{m}$, which corresponds to the same conditions as in the hierarchical system shown in Fig.~\ref{fig:fig3}.
After this initial stage of assembly is complete, we freeze the distribution of filaments and activate binding sites on the sides of the filaments, so that the filaments can aggregate vertically.
We assume that filaments are rigid and that no branching occurs, i.e. every filament may only bind to a single other filament above and below.
The total binding energy between two filaments then depends on their overlap; we assume that the vertical binding energy between two filaments of length $n_i$ and $n_j$ is given by
\begin{equation}
    \varepsilon_\mathrm{v} = \varepsilon_{\mathrm{v},0} \min(n_i, n_j) \,,
\end{equation}
where $\varepsilon_{\mathrm{v}, 0}$ is the per-particle vertical energy; we set $\varepsilon_{\mathrm{v}, 0} = 0.1~\mathrm{kT}$.
We neglect any entropic contributions resulting from the different possible binding offsets between two filaments, and assume that filaments are always overlapping without any ``overshoot'', resulting in $\min(n_i, n_j)$ particle contacts.

For only a single species of filament, the ``height'' distribution of the resulting sheets, $p(h)$, is monotonically decaying and is not well controllable, analogously to the single-species filament distribution, Eq.~\eqref{eq:nonaddressable}, as shown in Fig.~S2 in Supplementary Information.
To obtain more control over sheet heights, we introduce $m_2 = 10$ species of filaments, each consisting of $m_1=10$ distinct particle species, leading to a total species count of $m = 100$. (Even though $m=100$ species is pushing the boundaries of current experimental feasibility, the crucial point is that this does not scale with the size of the system, motivating an experimental challenge for the coming decade.)
In full analogy to the one-dimensional case shown in Fig.~\ref{fig:fig1}(c), each filament species can bind vertically to itself or to the subsequent filament species.
For simplicity, we further assume that all interactions during the final assembly step of assembly are infinitely strong, meaning that bonds no longer break and that each assembled sheet contains all 10 filament species.

Under these assumptions, we can obtain the size distributions for the multi-species sheets by sampling different sheets obtained from a single-species simulation and concatenating them together.
We start from a population of 16698 single-species sheets (the size distribution of which is shown in Fig.~S1), and then sample $10^6$ groups of 10 sheets each, which we concatenate vertically to simulate a multi-species sheet, leading to the size distribution as shown in Fig.~\ref{fig:fig4}.


\begin{thebibliography}{64}%
\makeatletter
\providecommand \@ifxundefined [1]{%
 \@ifx{#1\undefined}
}%
\providecommand \@ifnum [1]{%
 \ifnum #1\expandafter \@firstoftwo
 \else \expandafter \@secondoftwo
 \fi
}%
\providecommand \@ifx [1]{%
 \ifx #1\expandafter \@firstoftwo
 \else \expandafter \@secondoftwo
 \fi
}%
\providecommand \natexlab [1]{#1}%
\providecommand \enquote  [1]{``#1''}%
\providecommand \bibnamefont  [1]{#1}%
\providecommand \bibfnamefont [1]{#1}%
\providecommand \citenamefont [1]{#1}%
\providecommand \href@noop [0]{\@secondoftwo}%
\providecommand \href [0]{\begingroup \@sanitize@url \@href}%
\providecommand \@href[1]{\@@startlink{#1}\@@href}%
\providecommand \@@href[1]{\endgroup#1\@@endlink}%
\providecommand \@sanitize@url [0]{\catcode `\\12\catcode `\$12\catcode `\&12\catcode `\#12\catcode `\^12\catcode `\_12\catcode `\%12\relax}%
\providecommand \@@startlink[1]{}%
\providecommand \@@endlink[0]{}%
\providecommand \url  [0]{\begingroup\@sanitize@url \@url }%
\providecommand \@url [1]{\endgroup\@href {#1}{\urlprefix }}%
\providecommand \urlprefix  [0]{URL }%
\providecommand \Eprint [0]{\href }%
\providecommand \doibase [0]{https://doi.org/}%
\providecommand \selectlanguage [0]{\@gobble}%
\providecommand \bibinfo  [0]{\@secondoftwo}%
\providecommand \bibfield  [0]{\@secondoftwo}%
\providecommand \translation [1]{[#1]}%
\providecommand \BibitemOpen [0]{}%
\providecommand \bibitemStop [0]{}%
\providecommand \bibitemNoStop [0]{.\EOS\space}%
\providecommand \EOS [0]{\spacefactor3000\relax}%
\providecommand \BibitemShut  [1]{\csname bibitem#1\endcsname}%
\let\auto@bib@innerbib\@empty
\bibitem [{\citenamefont {Alberts}\ \emph {et~al.}(2015)\citenamefont {Alberts}, \citenamefont {Johnson}, \citenamefont {Lewis}, \citenamefont {Morgan}, \citenamefont {Raff}, \citenamefont {Roberts},\ and\ \citenamefont {Walter}}]{Alberts2015}%
  \BibitemOpen
  \bibfield  {author} {\bibinfo {author} {\bibfnamefont {B.}~\bibnamefont {Alberts}}, \bibinfo {author} {\bibfnamefont {A.}~\bibnamefont {Johnson}}, \bibinfo {author} {\bibfnamefont {J.}~\bibnamefont {Lewis}}, \bibinfo {author} {\bibfnamefont {D.}~\bibnamefont {Morgan}}, \bibinfo {author} {\bibfnamefont {M.}~\bibnamefont {Raff}}, \bibinfo {author} {\bibfnamefont {K.}~\bibnamefont {Roberts}},\ and\ \bibinfo {author} {\bibfnamefont {P.}~\bibnamefont {Walter}},\ }\href@noop {} {\emph {\bibinfo {title} {Molecular Biology of the Cell}}},\ \bibinfo {edition} {sixth}\ ed.\ (\bibinfo  {publisher} {Garland Science},\ \bibinfo {year} {2015})\BibitemShut {NoStop}%
\bibitem [{\citenamefont {Rhind}(2021)}]{Rhind.2021}%
  \BibitemOpen
  \bibfield  {author} {\bibinfo {author} {\bibfnamefont {N.}~\bibnamefont {Rhind}},\ }\bibfield  {title} {\bibinfo {title} {{Cell-size control}},\ }\href {https://doi.org/10.1016/j.cub.2021.09.017} {\bibfield  {journal} {\bibinfo  {journal} {Current Biology}\ }\textbf {\bibinfo {volume} {31}},\ \bibinfo {pages} {R1414} (\bibinfo {year} {2021})}\BibitemShut {NoStop}%
\bibitem [{\citenamefont {Ginzberg}\ \emph {et~al.}(2015)\citenamefont {Ginzberg}, \citenamefont {Kafri},\ and\ \citenamefont {Kirschner}}]{Ginzberg.2015}%
  \BibitemOpen
  \bibfield  {author} {\bibinfo {author} {\bibfnamefont {M.~B.}\ \bibnamefont {Ginzberg}}, \bibinfo {author} {\bibfnamefont {R.}~\bibnamefont {Kafri}},\ and\ \bibinfo {author} {\bibfnamefont {M.}~\bibnamefont {Kirschner}},\ }\bibfield  {title} {\bibinfo {title} {{On being the right (cell) size}},\ }\href {https://doi.org/10.1126/science.1245075} {\bibfield  {journal} {\bibinfo  {journal} {Science}\ }\textbf {\bibinfo {volume} {348}},\ \bibinfo {pages} {1245075} (\bibinfo {year} {2015})}\BibitemShut {NoStop}%
\bibitem [{\citenamefont {Hagan}\ and\ \citenamefont {Grason}(2021)}]{Hagan.2021}%
  \BibitemOpen
  \bibfield  {author} {\bibinfo {author} {\bibfnamefont {M.~F.}\ \bibnamefont {Hagan}}\ and\ \bibinfo {author} {\bibfnamefont {G.~M.}\ \bibnamefont {Grason}},\ }\bibfield  {title} {\bibinfo {title} {{Equilibrium mechanisms of self-limiting assembly}},\ }\href {https://doi.org/10.1103/revmodphys.93.025008} {\bibfield  {journal} {\bibinfo  {journal} {Reviews of Modern Physics}\ }\textbf {\bibinfo {volume} {93}},\ \bibinfo {pages} {025008} (\bibinfo {year} {2021})},\ \Eprint {https://arxiv.org/abs/2007.01927} {2007.01927} \BibitemShut {NoStop}%
\bibitem [{\citenamefont {Kim}\ \emph {et~al.}(2005)\citenamefont {Kim}, \citenamefont {Dalhaimer}, \citenamefont {Christian},\ and\ \citenamefont {Discher}}]{Kim.2005}%
  \BibitemOpen
  \bibfield  {author} {\bibinfo {author} {\bibfnamefont {Y.}~\bibnamefont {Kim}}, \bibinfo {author} {\bibfnamefont {P.}~\bibnamefont {Dalhaimer}}, \bibinfo {author} {\bibfnamefont {D.~A.}\ \bibnamefont {Christian}},\ and\ \bibinfo {author} {\bibfnamefont {D.~E.}\ \bibnamefont {Discher}},\ }\bibfield  {title} {\bibinfo {title} {{Polymeric worm micelles as nano-carriers for drug delivery}},\ }\href {https://doi.org/10.1088/0957-4484/16/7/024} {\bibfield  {journal} {\bibinfo  {journal} {Nanotechnology}\ }\textbf {\bibinfo {volume} {16}},\ \bibinfo {pages} {S484} (\bibinfo {year} {2005})}\BibitemShut {NoStop}%
\bibitem [{\citenamefont {Rad-Malekshahi}\ \emph {et~al.}(2016)\citenamefont {Rad-Malekshahi}, \citenamefont {Lempsink}, \citenamefont {Amidi}, \citenamefont {Hennink},\ and\ \citenamefont {Mastrobattista}}]{Rad-Malekshahi.2016}%
  \BibitemOpen
  \bibfield  {author} {\bibinfo {author} {\bibfnamefont {M.}~\bibnamefont {Rad-Malekshahi}}, \bibinfo {author} {\bibfnamefont {L.}~\bibnamefont {Lempsink}}, \bibinfo {author} {\bibfnamefont {M.}~\bibnamefont {Amidi}}, \bibinfo {author} {\bibfnamefont {W.~E.}\ \bibnamefont {Hennink}},\ and\ \bibinfo {author} {\bibfnamefont {E.}~\bibnamefont {Mastrobattista}},\ }\bibfield  {title} {\bibinfo {title} {{Biomedical Applications of Self-Assembling Peptides}},\ }\href {https://doi.org/10.1021/acs.bioconjchem.5b00487} {\bibfield  {journal} {\bibinfo  {journal} {Bioconjugate Chemistry}\ }\textbf {\bibinfo {volume} {27}},\ \bibinfo {pages} {3} (\bibinfo {year} {2016})}\BibitemShut {NoStop}%
\bibitem [{\citenamefont {Sigl}\ \emph {et~al.}(2021)\citenamefont {Sigl}, \citenamefont {Willner}, \citenamefont {Engelen}, \citenamefont {Kretzmann}, \citenamefont {Sachenbacher}, \citenamefont {Liedl}, \citenamefont {Kolbe}, \citenamefont {Wilsch}, \citenamefont {Aghvami}, \citenamefont {Protzer}, \citenamefont {Hagan}, \citenamefont {Fraden},\ and\ \citenamefont {Dietz}}]{Sigl.2021}%
  \BibitemOpen
  \bibfield  {author} {\bibinfo {author} {\bibfnamefont {C.}~\bibnamefont {Sigl}}, \bibinfo {author} {\bibfnamefont {E.~M.}\ \bibnamefont {Willner}}, \bibinfo {author} {\bibfnamefont {W.}~\bibnamefont {Engelen}}, \bibinfo {author} {\bibfnamefont {J.~A.}\ \bibnamefont {Kretzmann}}, \bibinfo {author} {\bibfnamefont {K.}~\bibnamefont {Sachenbacher}}, \bibinfo {author} {\bibfnamefont {A.}~\bibnamefont {Liedl}}, \bibinfo {author} {\bibfnamefont {F.}~\bibnamefont {Kolbe}}, \bibinfo {author} {\bibfnamefont {F.}~\bibnamefont {Wilsch}}, \bibinfo {author} {\bibfnamefont {S.~A.}\ \bibnamefont {Aghvami}}, \bibinfo {author} {\bibfnamefont {U.}~\bibnamefont {Protzer}}, \bibinfo {author} {\bibfnamefont {M.~F.}\ \bibnamefont {Hagan}}, \bibinfo {author} {\bibfnamefont {S.}~\bibnamefont {Fraden}},\ and\ \bibinfo {author} {\bibfnamefont {H.}~\bibnamefont {Dietz}},\ }\bibfield  {title} {\bibinfo {title} {{Programmable icosahedral shell system for virus trapping}},\ }\href {https://doi.org/10.1038/s41563-021-01020-4} {\bibfield
  {journal} {\bibinfo  {journal} {Nature Materials}\ }\textbf {\bibinfo {volume} {20}},\ \bibinfo {pages} {1281} (\bibinfo {year} {2021})}\BibitemShut {NoStop}%
\bibitem [{\citenamefont {Vukusic}\ and\ \citenamefont {Sambles}(2003)}]{Vukusic.2003}%
  \BibitemOpen
  \bibfield  {author} {\bibinfo {author} {\bibfnamefont {P.}~\bibnamefont {Vukusic}}\ and\ \bibinfo {author} {\bibfnamefont {J.~R.}\ \bibnamefont {Sambles}},\ }\bibfield  {title} {\bibinfo {title} {{Photonic structures in biology}},\ }\href {https://doi.org/10.1038/nature01941} {\bibfield  {journal} {\bibinfo  {journal} {Nature}\ }\textbf {\bibinfo {volume} {424}},\ \bibinfo {pages} {852} (\bibinfo {year} {2003})}\BibitemShut {NoStop}%
\bibitem [{\citenamefont {Dufresne}\ \emph {et~al.}(2009)\citenamefont {Dufresne}, \citenamefont {Noh}, \citenamefont {Saranathan}, \citenamefont {Mochrie}, \citenamefont {Cao},\ and\ \citenamefont {Prum}}]{Dufresne.2009}%
  \BibitemOpen
  \bibfield  {author} {\bibinfo {author} {\bibfnamefont {E.~R.}\ \bibnamefont {Dufresne}}, \bibinfo {author} {\bibfnamefont {H.}~\bibnamefont {Noh}}, \bibinfo {author} {\bibfnamefont {V.}~\bibnamefont {Saranathan}}, \bibinfo {author} {\bibfnamefont {S.~G.~J.}\ \bibnamefont {Mochrie}}, \bibinfo {author} {\bibfnamefont {H.}~\bibnamefont {Cao}},\ and\ \bibinfo {author} {\bibfnamefont {R.~O.}\ \bibnamefont {Prum}},\ }\bibfield  {title} {\bibinfo {title} {{Self-assembly of amorphous biophotonic nanostructures by phase separation}},\ }\href {https://doi.org/10.1039/b902775k} {\bibfield  {journal} {\bibinfo  {journal} {Soft Matter}\ }\textbf {\bibinfo {volume} {5}},\ \bibinfo {pages} {1792} (\bibinfo {year} {2009})}\BibitemShut {NoStop}%
\bibitem [{\citenamefont {Satyabola}\ \emph {et~al.}(2025)\citenamefont {Satyabola}, \citenamefont {Prasad}, \citenamefont {Yan},\ and\ \citenamefont {Zhou}}]{Satyabola.2025}%
  \BibitemOpen
  \bibfield  {author} {\bibinfo {author} {\bibfnamefont {D.}~\bibnamefont {Satyabola}}, \bibinfo {author} {\bibfnamefont {A.}~\bibnamefont {Prasad}}, \bibinfo {author} {\bibfnamefont {H.}~\bibnamefont {Yan}},\ and\ \bibinfo {author} {\bibfnamefont {X.}~\bibnamefont {Zhou}},\ }\bibfield  {title} {\bibinfo {title} {{Bioinspired Photonic Systems Directed by Designer DNA Nanostructures}},\ }\href {https://doi.org/10.1021/acsaom.4c00103} {\bibfield  {journal} {\bibinfo  {journal} {ACS Applied Optical Materials}\ }\textbf {\bibinfo {volume} {3}},\ \bibinfo {pages} {552} (\bibinfo {year} {2025})}\BibitemShut {NoStop}%
\bibitem [{\citenamefont {Hayakawa}\ \emph {et~al.}(2024)\citenamefont {Hayakawa}, \citenamefont {Videbæk}, \citenamefont {Grason},\ and\ \citenamefont {Rogers}}]{Hayakawa.2024}%
  \BibitemOpen
  \bibfield  {author} {\bibinfo {author} {\bibfnamefont {D.}~\bibnamefont {Hayakawa}}, \bibinfo {author} {\bibfnamefont {T.~E.}\ \bibnamefont {Videbæk}}, \bibinfo {author} {\bibfnamefont {G.~M.}\ \bibnamefont {Grason}},\ and\ \bibinfo {author} {\bibfnamefont {W.~B.}\ \bibnamefont {Rogers}},\ }\bibfield  {title} {\bibinfo {title} {{Symmetry-Guided Inverse Design of Self-Assembling Multiscale DNA Origami Tilings}},\ }\href {https://doi.org/10.1021/acsnano.4c04515} {\bibfield  {journal} {\bibinfo  {journal} {ACS Nano}\ }\textbf {\bibinfo {volume} {18}},\ \bibinfo {pages} {19169} (\bibinfo {year} {2024})}\BibitemShut {NoStop}%
\bibitem [{\citenamefont {Hensley}\ \emph {et~al.}(2023)\citenamefont {Hensley}, \citenamefont {Videbæk}, \citenamefont {Seyforth}, \citenamefont {Jacobs},\ and\ \citenamefont {Rogers}}]{Hensley.2023}%
  \BibitemOpen
  \bibfield  {author} {\bibinfo {author} {\bibfnamefont {A.}~\bibnamefont {Hensley}}, \bibinfo {author} {\bibfnamefont {T.~E.}\ \bibnamefont {Videbæk}}, \bibinfo {author} {\bibfnamefont {H.}~\bibnamefont {Seyforth}}, \bibinfo {author} {\bibfnamefont {W.~M.}\ \bibnamefont {Jacobs}},\ and\ \bibinfo {author} {\bibfnamefont {W.~B.}\ \bibnamefont {Rogers}},\ }\bibfield  {title} {\bibinfo {title} {{Macroscopic photonic single crystals via seeded growth of DNA-coated colloids}},\ }\href {https://doi.org/10.1038/s41467-023-39992-3} {\bibfield  {journal} {\bibinfo  {journal} {Nature Communications}\ }\textbf {\bibinfo {volume} {14}},\ \bibinfo {pages} {4237} (\bibinfo {year} {2023})},\ \Eprint {https://arxiv.org/abs/2303.04074} {2303.04074} \BibitemShut {NoStop}%
\bibitem [{\citenamefont {Aldaye}\ \emph {et~al.}(2008)\citenamefont {Aldaye}, \citenamefont {Palmer},\ and\ \citenamefont {Sleiman}}]{Aldaye.2008}%
  \BibitemOpen
  \bibfield  {author} {\bibinfo {author} {\bibfnamefont {F.~A.}\ \bibnamefont {Aldaye}}, \bibinfo {author} {\bibfnamefont {A.~L.}\ \bibnamefont {Palmer}},\ and\ \bibinfo {author} {\bibfnamefont {H.~F.}\ \bibnamefont {Sleiman}},\ }\bibfield  {title} {\bibinfo {title} {{Assembling Materials with DNA as the Guide}},\ }\href {https://doi.org/10.1126/science.1154533} {\bibfield  {journal} {\bibinfo  {journal} {Science}\ }\textbf {\bibinfo {volume} {321}},\ \bibinfo {pages} {1795} (\bibinfo {year} {2008})}\BibitemShut {NoStop}%
\bibitem [{\citenamefont {Pearce}\ \emph {et~al.}(2021)\citenamefont {Pearce}, \citenamefont {Wilks}, \citenamefont {Arno},\ and\ \citenamefont {O’Reilly}}]{Pearce.2021}%
  \BibitemOpen
  \bibfield  {author} {\bibinfo {author} {\bibfnamefont {A.~K.}\ \bibnamefont {Pearce}}, \bibinfo {author} {\bibfnamefont {T.~R.}\ \bibnamefont {Wilks}}, \bibinfo {author} {\bibfnamefont {M.~C.}\ \bibnamefont {Arno}},\ and\ \bibinfo {author} {\bibfnamefont {R.~K.}\ \bibnamefont {O’Reilly}},\ }\bibfield  {title} {\bibinfo {title} {{Synthesis and applications of anisotropic nanoparticles with precisely defined dimensions}},\ }\href {https://doi.org/10.1038/s41570-020-00232-7} {\bibfield  {journal} {\bibinfo  {journal} {Nature Reviews Chemistry}\ }\textbf {\bibinfo {volume} {5}},\ \bibinfo {pages} {21} (\bibinfo {year} {2021})}\BibitemShut {NoStop}%
\bibitem [{\citenamefont {Michelson}\ \emph {et~al.}(2024)\citenamefont {Michelson}, \citenamefont {Subramanian}, \citenamefont {Kisslinger}, \citenamefont {Tiwale}, \citenamefont {Xiang}, \citenamefont {Shen}, \citenamefont {Kahn}, \citenamefont {Nykypanchuk}, \citenamefont {Yan}, \citenamefont {Nam},\ and\ \citenamefont {Gang}}]{Michelson.2024}%
  \BibitemOpen
  \bibfield  {author} {\bibinfo {author} {\bibfnamefont {A.}~\bibnamefont {Michelson}}, \bibinfo {author} {\bibfnamefont {A.}~\bibnamefont {Subramanian}}, \bibinfo {author} {\bibfnamefont {K.}~\bibnamefont {Kisslinger}}, \bibinfo {author} {\bibfnamefont {N.}~\bibnamefont {Tiwale}}, \bibinfo {author} {\bibfnamefont {S.}~\bibnamefont {Xiang}}, \bibinfo {author} {\bibfnamefont {E.}~\bibnamefont {Shen}}, \bibinfo {author} {\bibfnamefont {J.~S.}\ \bibnamefont {Kahn}}, \bibinfo {author} {\bibfnamefont {D.}~\bibnamefont {Nykypanchuk}}, \bibinfo {author} {\bibfnamefont {H.}~\bibnamefont {Yan}}, \bibinfo {author} {\bibfnamefont {C.-Y.}\ \bibnamefont {Nam}},\ and\ \bibinfo {author} {\bibfnamefont {O.}~\bibnamefont {Gang}},\ }\bibfield  {title} {\bibinfo {title} {{Three-dimensional nanoscale metal, metal oxide, and semiconductor frameworks through DNA-programmable assembly and templating}},\ }\href {https://doi.org/10.1126/sciadv.adl0604} {\bibfield  {journal} {\bibinfo  {journal} {Science Advances}\ }\textbf {\bibinfo
  {volume} {10}},\ \bibinfo {pages} {eadl0604} (\bibinfo {year} {2024})}\BibitemShut {NoStop}%
\bibitem [{\citenamefont {Michelson}\ \emph {et~al.}(2025)\citenamefont {Michelson}, \citenamefont {Shani}, \citenamefont {Kahn}, \citenamefont {Redeker}, \citenamefont {Lee}, \citenamefont {DeOlivares}, \citenamefont {Kisslinger}, \citenamefont {Tiwale}, \citenamefont {Yan}, \citenamefont {Pattammattel}, \citenamefont {Nam}, \citenamefont {Pribiag},\ and\ \citenamefont {Gang}}]{Michelson.2025}%
  \BibitemOpen
  \bibfield  {author} {\bibinfo {author} {\bibfnamefont {A.}~\bibnamefont {Michelson}}, \bibinfo {author} {\bibfnamefont {L.}~\bibnamefont {Shani}}, \bibinfo {author} {\bibfnamefont {J.~S.}\ \bibnamefont {Kahn}}, \bibinfo {author} {\bibfnamefont {D.~C.}\ \bibnamefont {Redeker}}, \bibinfo {author} {\bibfnamefont {W.-I.}\ \bibnamefont {Lee}}, \bibinfo {author} {\bibfnamefont {K.~R.}\ \bibnamefont {DeOlivares}}, \bibinfo {author} {\bibfnamefont {K.}~\bibnamefont {Kisslinger}}, \bibinfo {author} {\bibfnamefont {N.}~\bibnamefont {Tiwale}}, \bibinfo {author} {\bibfnamefont {H.}~\bibnamefont {Yan}}, \bibinfo {author} {\bibfnamefont {A.}~\bibnamefont {Pattammattel}}, \bibinfo {author} {\bibfnamefont {C.-Y.}\ \bibnamefont {Nam}}, \bibinfo {author} {\bibfnamefont {V.~S.}\ \bibnamefont {Pribiag}},\ and\ \bibinfo {author} {\bibfnamefont {O.}~\bibnamefont {Gang}},\ }\bibfield  {title} {\bibinfo {title} {{Scalable fabrication of Chip-integrated 3D-nanostructured electronic devices via DNA-programmable assembly}},\ }\href
  {https://doi.org/10.1126/sciadv.adt5620} {\bibfield  {journal} {\bibinfo  {journal} {Science Advances}\ }\textbf {\bibinfo {volume} {11}},\ \bibinfo {pages} {eadt5620} (\bibinfo {year} {2025})}\BibitemShut {NoStop}%
\bibitem [{\citenamefont {Hayakawa}\ \emph {et~al.}(2022)\citenamefont {Hayakawa}, \citenamefont {Videbaek}, \citenamefont {Hall}, \citenamefont {Fang}, \citenamefont {Sigl}, \citenamefont {Feigl}, \citenamefont {Dietz}, \citenamefont {Fraden}, \citenamefont {Hagan}, \citenamefont {Grason},\ and\ \citenamefont {Rogers}}]{Hayakawa.2022pqs}%
  \BibitemOpen
  \bibfield  {author} {\bibinfo {author} {\bibfnamefont {D.}~\bibnamefont {Hayakawa}}, \bibinfo {author} {\bibfnamefont {T.~E.}\ \bibnamefont {Videbaek}}, \bibinfo {author} {\bibfnamefont {D.~M.}\ \bibnamefont {Hall}}, \bibinfo {author} {\bibfnamefont {H.}~\bibnamefont {Fang}}, \bibinfo {author} {\bibfnamefont {C.}~\bibnamefont {Sigl}}, \bibinfo {author} {\bibfnamefont {E.}~\bibnamefont {Feigl}}, \bibinfo {author} {\bibfnamefont {H.}~\bibnamefont {Dietz}}, \bibinfo {author} {\bibfnamefont {S.}~\bibnamefont {Fraden}}, \bibinfo {author} {\bibfnamefont {M.~F.}\ \bibnamefont {Hagan}}, \bibinfo {author} {\bibfnamefont {G.~M.}\ \bibnamefont {Grason}},\ and\ \bibinfo {author} {\bibfnamefont {W.~B.}\ \bibnamefont {Rogers}},\ }\bibfield  {title} {\bibinfo {title} {{Geometrically programmed self-limited assembly of tubules using DNA origami colloids}},\ }\href {https://doi.org/10.1073/pnas.2207902119} {\bibfield  {journal} {\bibinfo  {journal} {Proceedings of the National Academy of Sciences}\ }\textbf {\bibinfo
  {volume} {119}},\ \bibinfo {pages} {e2207902119} (\bibinfo {year} {2022})},\ \Eprint {https://arxiv.org/abs/2203.01421} {2203.01421} \BibitemShut {NoStop}%
\bibitem [{\citenamefont {Videbæk}\ \emph {et~al.}(2024)\citenamefont {Videbæk}, \citenamefont {Hayakawa}, \citenamefont {Grason}, \citenamefont {Hagan}, \citenamefont {Fraden},\ and\ \citenamefont {Rogers}}]{Videbaek.2024}%
  \BibitemOpen
  \bibfield  {author} {\bibinfo {author} {\bibfnamefont {T.~E.}\ \bibnamefont {Videbæk}}, \bibinfo {author} {\bibfnamefont {D.}~\bibnamefont {Hayakawa}}, \bibinfo {author} {\bibfnamefont {G.~M.}\ \bibnamefont {Grason}}, \bibinfo {author} {\bibfnamefont {M.~F.}\ \bibnamefont {Hagan}}, \bibinfo {author} {\bibfnamefont {S.}~\bibnamefont {Fraden}},\ and\ \bibinfo {author} {\bibfnamefont {W.~B.}\ \bibnamefont {Rogers}},\ }\bibfield  {title} {\bibinfo {title} {{Economical routes to size-specific assembly of self-closing structures}},\ }\href {https://doi.org/10.1126/sciadv.ado5979} {\bibfield  {journal} {\bibinfo  {journal} {Science Advances}\ }\textbf {\bibinfo {volume} {10}},\ \bibinfo {pages} {eado5979} (\bibinfo {year} {2024})}\BibitemShut {NoStop}%
\bibitem [{\citenamefont {Duque}\ \emph {et~al.}(2024)\citenamefont {Duque}, \citenamefont {Hall}, \citenamefont {Tyukodi}, \citenamefont {Hagan}, \citenamefont {Santangelo},\ and\ \citenamefont {Grason}}]{Duque.2024}%
  \BibitemOpen
  \bibfield  {author} {\bibinfo {author} {\bibfnamefont {C.~M.}\ \bibnamefont {Duque}}, \bibinfo {author} {\bibfnamefont {D.~M.}\ \bibnamefont {Hall}}, \bibinfo {author} {\bibfnamefont {B.}~\bibnamefont {Tyukodi}}, \bibinfo {author} {\bibfnamefont {M.~F.}\ \bibnamefont {Hagan}}, \bibinfo {author} {\bibfnamefont {C.~D.}\ \bibnamefont {Santangelo}},\ and\ \bibinfo {author} {\bibfnamefont {G.~M.}\ \bibnamefont {Grason}},\ }\bibfield  {title} {\bibinfo {title} {{Limits of economy and fidelity for programmable assembly of size-controlled triply periodic polyhedra}},\ }\href {https://doi.org/10.1073/pnas.2315648121} {\bibfield  {journal} {\bibinfo  {journal} {Proceedings of the National Academy of Sciences}\ }\textbf {\bibinfo {volume} {121}},\ \bibinfo {pages} {e2315648121} (\bibinfo {year} {2024})}\BibitemShut {NoStop}%
\bibitem [{\citenamefont {Saha}\ \emph {et~al.}(2025)\citenamefont {Saha}, \citenamefont {Hayakawa}, \citenamefont {Videbaek}, \citenamefont {Price}, \citenamefont {Wei}, \citenamefont {Pombo}, \citenamefont {Duke}, \citenamefont {Arya}, \citenamefont {Grason}, \citenamefont {Rogers},\ and\ \citenamefont {Fraden}}]{Saha.2025}%
  \BibitemOpen
  \bibfield  {author} {\bibinfo {author} {\bibfnamefont {R.}~\bibnamefont {Saha}}, \bibinfo {author} {\bibfnamefont {D.}~\bibnamefont {Hayakawa}}, \bibinfo {author} {\bibfnamefont {T.~E.}\ \bibnamefont {Videbaek}}, \bibinfo {author} {\bibfnamefont {M.}~\bibnamefont {Price}}, \bibinfo {author} {\bibfnamefont {W.-S.}\ \bibnamefont {Wei}}, \bibinfo {author} {\bibfnamefont {J.}~\bibnamefont {Pombo}}, \bibinfo {author} {\bibfnamefont {D.}~\bibnamefont {Duke}}, \bibinfo {author} {\bibfnamefont {G.}~\bibnamefont {Arya}}, \bibinfo {author} {\bibfnamefont {G.~M.}\ \bibnamefont {Grason}}, \bibinfo {author} {\bibfnamefont {W.~B.}\ \bibnamefont {Rogers}},\ and\ \bibinfo {author} {\bibfnamefont {S.}~\bibnamefont {Fraden}},\ }\bibfield  {title} {\bibinfo {title} {{Modular programming of interaction and geometric specificity enables assembly of complex DNA origami nanostructures}},\ }\href@noop {} {\bibfield  {journal} {\bibinfo  {journal} {arXiv}\ } (\bibinfo {year} {2025})},\ \Eprint {https://arxiv.org/abs/2502.05388}
  {2502.05388} \BibitemShut {NoStop}%
\bibitem [{\citenamefont {Sciortino}\ \emph {et~al.}(2004)\citenamefont {Sciortino}, \citenamefont {Mossa}, \citenamefont {Zaccarelli},\ and\ \citenamefont {Tartaglia}}]{Sciortino.2004}%
  \BibitemOpen
  \bibfield  {author} {\bibinfo {author} {\bibfnamefont {F.}~\bibnamefont {Sciortino}}, \bibinfo {author} {\bibfnamefont {S.}~\bibnamefont {Mossa}}, \bibinfo {author} {\bibfnamefont {E.}~\bibnamefont {Zaccarelli}},\ and\ \bibinfo {author} {\bibfnamefont {P.}~\bibnamefont {Tartaglia}},\ }\bibfield  {title} {\bibinfo {title} {{Equilibrium Cluster Phases and Low-Density Arrested Disordered States: The Role of Short-Range Attraction and Long-Range Repulsion}},\ }\href {https://doi.org/10.1103/physrevlett.93.055701} {\bibfield  {journal} {\bibinfo  {journal} {Physical Review Letters}\ }\textbf {\bibinfo {volume} {93}},\ \bibinfo {pages} {055701} (\bibinfo {year} {2004})},\ \Eprint {https://arxiv.org/abs/cond-mat/0312161} {cond-mat/0312161} \BibitemShut {NoStop}%
\bibitem [{\citenamefont {Nguyen}\ \emph {et~al.}(2015)\citenamefont {Nguyen}, \citenamefont {Schultz}, \citenamefont {Kotov},\ and\ \citenamefont {Glotzer}}]{Nguyen.2015}%
  \BibitemOpen
  \bibfield  {author} {\bibinfo {author} {\bibfnamefont {T.~D.}\ \bibnamefont {Nguyen}}, \bibinfo {author} {\bibfnamefont {B.~A.}\ \bibnamefont {Schultz}}, \bibinfo {author} {\bibfnamefont {N.~A.}\ \bibnamefont {Kotov}},\ and\ \bibinfo {author} {\bibfnamefont {S.~C.}\ \bibnamefont {Glotzer}},\ }\bibfield  {title} {\bibinfo {title} {{Generic, phenomenological, on-the-fly renormalized repulsion model for self-limited organization of terminal supraparticle assemblies}},\ }\href {https://doi.org/10.1073/pnas.1509239112} {\bibfield  {journal} {\bibinfo  {journal} {Proceedings of the National Academy of Sciences}\ }\textbf {\bibinfo {volume} {112}},\ \bibinfo {pages} {E3161} (\bibinfo {year} {2015})}\BibitemShut {NoStop}%
\bibitem [{\citenamefont {Grason}(2016)}]{Grason.2016}%
  \BibitemOpen
  \bibfield  {author} {\bibinfo {author} {\bibfnamefont {G.~M.}\ \bibnamefont {Grason}},\ }\bibfield  {title} {\bibinfo {title} {{Perspective: Geometrically frustrated assemblies}},\ }\href {https://doi.org/10.1063/1.4962629} {\bibfield  {journal} {\bibinfo  {journal} {The Journal of Chemical Physics}\ }\textbf {\bibinfo {volume} {145}},\ \bibinfo {pages} {110901} (\bibinfo {year} {2016})},\ \Eprint {https://arxiv.org/abs/1608.07833} {1608.07833} \BibitemShut {NoStop}%
\bibitem [{\citenamefont {Lenz}\ and\ \citenamefont {Witten}(2017)}]{lenz2017geometrical-1f1}%
  \BibitemOpen
  \bibfield  {author} {\bibinfo {author} {\bibfnamefont {M.}~\bibnamefont {Lenz}}\ and\ \bibinfo {author} {\bibfnamefont {T.~A.}\ \bibnamefont {Witten}},\ }\bibfield  {title} {\bibinfo {title} {Geometrical frustration yields fibre formation in self-assembly},\ }\href {https://doi.org/10.1038/nphys4184} {\bibfield  {journal} {\bibinfo  {journal} {Nature Physics}\ }\textbf {\bibinfo {volume} {13}},\ \bibinfo {pages} {1100} (\bibinfo {year} {2017})},\ \Eprint {https://arxiv.org/abs/1705.08334} {1705.08334} \BibitemShut {NoStop}%
\bibitem [{\citenamefont {Roy}\ \emph {et~al.}(2025)\citenamefont {Roy}, \citenamefont {Terzi},\ and\ \citenamefont {Lenz}}]{roy2025collective-bd3}%
  \BibitemOpen
  \bibfield  {author} {\bibinfo {author} {\bibfnamefont {H.~L.}\ \bibnamefont {Roy}}, \bibinfo {author} {\bibfnamefont {M.~M.}\ \bibnamefont {Terzi}},\ and\ \bibinfo {author} {\bibfnamefont {M.}~\bibnamefont {Lenz}},\ }\bibfield  {title} {\bibinfo {title} {Collective deformation modes promote fibrous self-assembly in deformable particles},\ }\href {https://doi.org/10.1103/physrevx.15.011022} {\bibfield  {journal} {\bibinfo  {journal} {Physical Review X}\ }\textbf {\bibinfo {volume} {15}},\ \bibinfo {pages} {011022} (\bibinfo {year} {2025})}\BibitemShut {NoStop}%
\bibitem [{\citenamefont {Koehler}\ \emph {et~al.}(2025)\citenamefont {Koehler}, \citenamefont {Eder}, \citenamefont {Karfusehr}, \citenamefont {Ouazan-Reboul}, \citenamefont {Ronceray}, \citenamefont {Simmel},\ and\ \citenamefont {Lenz}}]{koehler2025topological-bb2}%
  \BibitemOpen
  \bibfield  {author} {\bibinfo {author} {\bibfnamefont {L.}~\bibnamefont {Koehler}}, \bibinfo {author} {\bibfnamefont {M.}~\bibnamefont {Eder}}, \bibinfo {author} {\bibfnamefont {C.}~\bibnamefont {Karfusehr}}, \bibinfo {author} {\bibfnamefont {V.}~\bibnamefont {Ouazan-Reboul}}, \bibinfo {author} {\bibfnamefont {P.}~\bibnamefont {Ronceray}}, \bibinfo {author} {\bibfnamefont {F.~C.}\ \bibnamefont {Simmel}},\ and\ \bibinfo {author} {\bibfnamefont {M.}~\bibnamefont {Lenz}},\ }\bibfield  {title} {\bibinfo {title} {Topological defect engineering enables size and shape control in self-assembly},\ }\bibfield  {journal} {\bibinfo  {journal} {{arXiv}}\ }\href {https://doi.org/10.48550/arxiv.2504.13073} {10.48550/arxiv.2504.13073} (\bibinfo {year} {2025}),\ \Eprint {https://arxiv.org/abs/2504.13073} {2504.13073} \BibitemShut {NoStop}%
\bibitem [{\citenamefont {Johann}\ \emph {et~al.}(2012)\citenamefont {Johann}, \citenamefont {Erlenkämper},\ and\ \citenamefont {Kruse}}]{Johann.2012}%
  \BibitemOpen
  \bibfield  {author} {\bibinfo {author} {\bibfnamefont {D.}~\bibnamefont {Johann}}, \bibinfo {author} {\bibfnamefont {C.}~\bibnamefont {Erlenkämper}},\ and\ \bibinfo {author} {\bibfnamefont {K.}~\bibnamefont {Kruse}},\ }\bibfield  {title} {\bibinfo {title} {{Length Regulation of Active Biopolymers by Molecular Motors}},\ }\href {https://doi.org/10.1103/physrevlett.108.258103} {\bibfield  {journal} {\bibinfo  {journal} {Physical Review Letters}\ }\textbf {\bibinfo {volume} {108}},\ \bibinfo {pages} {258103} (\bibinfo {year} {2012})}\BibitemShut {NoStop}%
\bibitem [{\citenamefont {Melbinger}\ \emph {et~al.}(2012)\citenamefont {Melbinger}, \citenamefont {Reese},\ and\ \citenamefont {Frey}}]{Melbinger.2012}%
  \BibitemOpen
  \bibfield  {author} {\bibinfo {author} {\bibfnamefont {A.}~\bibnamefont {Melbinger}}, \bibinfo {author} {\bibfnamefont {L.}~\bibnamefont {Reese}},\ and\ \bibinfo {author} {\bibfnamefont {E.}~\bibnamefont {Frey}},\ }\bibfield  {title} {\bibinfo {title} {{Microtubule Length Regulation by Molecular Motors}},\ }\href {https://doi.org/10.1103/physrevlett.108.258104} {\bibfield  {journal} {\bibinfo  {journal} {Physical Review Letters}\ }\textbf {\bibinfo {volume} {108}},\ \bibinfo {pages} {258104} (\bibinfo {year} {2012})},\ \Eprint {https://arxiv.org/abs/1204.5655} {1204.5655} \BibitemShut {NoStop}%
\bibitem [{\citenamefont {Striebel}\ \emph {et~al.}(2022)\citenamefont {Striebel}, \citenamefont {Brauns},\ and\ \citenamefont {Frey}}]{Striebel.2022}%
  \BibitemOpen
  \bibfield  {author} {\bibinfo {author} {\bibfnamefont {M.}~\bibnamefont {Striebel}}, \bibinfo {author} {\bibfnamefont {F.}~\bibnamefont {Brauns}},\ and\ \bibinfo {author} {\bibfnamefont {E.}~\bibnamefont {Frey}},\ }\bibfield  {title} {\bibinfo {title} {{Length Regulation Drives Self-Organization in Filament-Motor Mixtures}},\ }\href {https://doi.org/10.1103/physrevlett.129.238102} {\bibfield  {journal} {\bibinfo  {journal} {Physical Review Letters}\ }\textbf {\bibinfo {volume} {129}},\ \bibinfo {pages} {238102} (\bibinfo {year} {2022})},\ \Eprint {https://arxiv.org/abs/2109.05091} {2109.05091} \BibitemShut {NoStop}%
\bibitem [{\citenamefont {Kuan}\ and\ \citenamefont {Betterton}(2013)}]{Kuan.2013}%
  \BibitemOpen
  \bibfield  {author} {\bibinfo {author} {\bibfnamefont {H.-S.}\ \bibnamefont {Kuan}}\ and\ \bibinfo {author} {\bibfnamefont {M.~D.}\ \bibnamefont {Betterton}},\ }\bibfield  {title} {\bibinfo {title} {{Biophysics of filament length regulation by molecular motors}},\ }\href {https://doi.org/10.1088/1478-3975/10/3/036004} {\bibfield  {journal} {\bibinfo  {journal} {Physical Biology}\ }\textbf {\bibinfo {volume} {10}},\ \bibinfo {pages} {036004} (\bibinfo {year} {2013})},\ \Eprint {https://arxiv.org/abs/1302.3196} {1302.3196} \BibitemShut {NoStop}%
\bibitem [{\citenamefont {Datta}\ \emph {et~al.}(2018)\citenamefont {Datta}, \citenamefont {Saha},\ and\ \citenamefont {Stang}}]{Datta.2018}%
  \BibitemOpen
  \bibfield  {author} {\bibinfo {author} {\bibfnamefont {S.}~\bibnamefont {Datta}}, \bibinfo {author} {\bibfnamefont {M.~L.}\ \bibnamefont {Saha}},\ and\ \bibinfo {author} {\bibfnamefont {P.~J.}\ \bibnamefont {Stang}},\ }\bibfield  {title} {\bibinfo {title} {{Hierarchical Assemblies of Supramolecular Coordination Complexes}},\ }\href {https://doi.org/10.1021/acs.accounts.8b00233} {\bibfield  {journal} {\bibinfo  {journal} {Accounts of Chemical Research}\ }\textbf {\bibinfo {volume} {51}},\ \bibinfo {pages} {2047} (\bibinfo {year} {2018})}\BibitemShut {NoStop}%
\bibitem [{\citenamefont {Murugan}\ \emph {et~al.}(2015)\citenamefont {Murugan}, \citenamefont {Zou},\ and\ \citenamefont {Brenner}}]{Murugan.2015}%
  \BibitemOpen
  \bibfield  {author} {\bibinfo {author} {\bibfnamefont {A.}~\bibnamefont {Murugan}}, \bibinfo {author} {\bibfnamefont {J.}~\bibnamefont {Zou}},\ and\ \bibinfo {author} {\bibfnamefont {M.~P.}\ \bibnamefont {Brenner}},\ }\bibfield  {title} {\bibinfo {title} {{Undesired usage and the robust self-assembly of heterogeneous structures}},\ }\href {https://doi.org/10.1038/ncomms7203} {\bibfield  {journal} {\bibinfo  {journal} {Nature Communications}\ }\textbf {\bibinfo {volume} {6}},\ \bibinfo {pages} {6203} (\bibinfo {year} {2015})}\BibitemShut {NoStop}%
\bibitem [{\citenamefont {He}\ \emph {et~al.}(2020)\citenamefont {He}, \citenamefont {Gales}, \citenamefont {Ducrot}, \citenamefont {Gong}, \citenamefont {Yi}, \citenamefont {Sacanna},\ and\ \citenamefont {Pine}}]{He.2020}%
  \BibitemOpen
  \bibfield  {author} {\bibinfo {author} {\bibfnamefont {M.}~\bibnamefont {He}}, \bibinfo {author} {\bibfnamefont {J.~P.}\ \bibnamefont {Gales}}, \bibinfo {author} {\bibfnamefont {E.}~\bibnamefont {Ducrot}}, \bibinfo {author} {\bibfnamefont {Z.}~\bibnamefont {Gong}}, \bibinfo {author} {\bibfnamefont {G.-R.}\ \bibnamefont {Yi}}, \bibinfo {author} {\bibfnamefont {S.}~\bibnamefont {Sacanna}},\ and\ \bibinfo {author} {\bibfnamefont {D.~J.}\ \bibnamefont {Pine}},\ }\bibfield  {title} {\bibinfo {title} {{Colloidal diamond}},\ }\href {https://doi.org/10.1038/s41586-020-2718-6} {\bibfield  {journal} {\bibinfo  {journal} {Nature}\ }\textbf {\bibinfo {volume} {585}},\ \bibinfo {pages} {524} (\bibinfo {year} {2020})}\BibitemShut {NoStop}%
\bibitem [{\citenamefont {Mirkin}\ \emph {et~al.}(1996)\citenamefont {Mirkin}, \citenamefont {Letsinger}, \citenamefont {Mucic},\ and\ \citenamefont {Storhoff}}]{Mirkin.1996}%
  \BibitemOpen
  \bibfield  {author} {\bibinfo {author} {\bibfnamefont {C.~A.}\ \bibnamefont {Mirkin}}, \bibinfo {author} {\bibfnamefont {R.~L.}\ \bibnamefont {Letsinger}}, \bibinfo {author} {\bibfnamefont {R.~C.}\ \bibnamefont {Mucic}},\ and\ \bibinfo {author} {\bibfnamefont {J.~J.}\ \bibnamefont {Storhoff}},\ }\bibfield  {title} {\bibinfo {title} {{A DNA-based method for rationally assembling nanoparticles into macroscopic materials}},\ }\href {https://doi.org/10.1038/382607a0} {\bibfield  {journal} {\bibinfo  {journal} {Nature}\ }\textbf {\bibinfo {volume} {382}},\ \bibinfo {pages} {607} (\bibinfo {year} {1996})}\BibitemShut {NoStop}%
\bibitem [{\citenamefont {Rogers}\ and\ \citenamefont {Manoharan}(2015)}]{Rogers.2015}%
  \BibitemOpen
  \bibfield  {author} {\bibinfo {author} {\bibfnamefont {W.~B.}\ \bibnamefont {Rogers}}\ and\ \bibinfo {author} {\bibfnamefont {V.~N.}\ \bibnamefont {Manoharan}},\ }\bibfield  {title} {\bibinfo {title} {{Programming colloidal phase transitions with DNA strand displacement}},\ }\href {https://doi.org/10.1126/science.1259762} {\bibfield  {journal} {\bibinfo  {journal} {Science}\ }\textbf {\bibinfo {volume} {347}},\ \bibinfo {pages} {639} (\bibinfo {year} {2015})}\BibitemShut {NoStop}%
\bibitem [{\citenamefont {Wang}\ \emph {et~al.}(2012)\citenamefont {Wang}, \citenamefont {Wang}, \citenamefont {Breed}, \citenamefont {Manoharan}, \citenamefont {Feng}, \citenamefont {Hollingsworth}, \citenamefont {Weck},\ and\ \citenamefont {Pine}}]{Wang.2012}%
  \BibitemOpen
  \bibfield  {author} {\bibinfo {author} {\bibfnamefont {Y.}~\bibnamefont {Wang}}, \bibinfo {author} {\bibfnamefont {Y.}~\bibnamefont {Wang}}, \bibinfo {author} {\bibfnamefont {D.~R.}\ \bibnamefont {Breed}}, \bibinfo {author} {\bibfnamefont {V.~N.}\ \bibnamefont {Manoharan}}, \bibinfo {author} {\bibfnamefont {L.}~\bibnamefont {Feng}}, \bibinfo {author} {\bibfnamefont {A.~D.}\ \bibnamefont {Hollingsworth}}, \bibinfo {author} {\bibfnamefont {M.}~\bibnamefont {Weck}},\ and\ \bibinfo {author} {\bibfnamefont {D.~J.}\ \bibnamefont {Pine}},\ }\bibfield  {title} {\bibinfo {title} {{Colloids with valence and specific directional bonding}},\ }\href {https://doi.org/10.1038/nature11564} {\bibfield  {journal} {\bibinfo  {journal} {Nature}\ }\textbf {\bibinfo {volume} {491}},\ \bibinfo {pages} {51} (\bibinfo {year} {2012})}\BibitemShut {NoStop}%
\bibitem [{\citenamefont {Wang}\ \emph {et~al.}(2015)\citenamefont {Wang}, \citenamefont {Wang}, \citenamefont {Zheng}, \citenamefont {Ducrot}, \citenamefont {Yodh}, \citenamefont {Weck},\ and\ \citenamefont {Pine}}]{Wang.2015}%
  \BibitemOpen
  \bibfield  {author} {\bibinfo {author} {\bibfnamefont {Y.}~\bibnamefont {Wang}}, \bibinfo {author} {\bibfnamefont {Y.}~\bibnamefont {Wang}}, \bibinfo {author} {\bibfnamefont {X.}~\bibnamefont {Zheng}}, \bibinfo {author} {\bibfnamefont {E.}~\bibnamefont {Ducrot}}, \bibinfo {author} {\bibfnamefont {J.~S.}\ \bibnamefont {Yodh}}, \bibinfo {author} {\bibfnamefont {M.}~\bibnamefont {Weck}},\ and\ \bibinfo {author} {\bibfnamefont {D.~J.}\ \bibnamefont {Pine}},\ }\bibfield  {title} {\bibinfo {title} {{Crystallization of DNA-coated colloids}},\ }\href {https://doi.org/10.1038/ncomms8253} {\bibfield  {journal} {\bibinfo  {journal} {Nature Communications}\ }\textbf {\bibinfo {volume} {6}},\ \bibinfo {pages} {7253} (\bibinfo {year} {2015})}\BibitemShut {NoStop}%
\bibitem [{\citenamefont {Valignat}\ \emph {et~al.}(2005)\citenamefont {Valignat}, \citenamefont {Theodoly}, \citenamefont {Crocker}, \citenamefont {Russel},\ and\ \citenamefont {Chaikin}}]{Valignat.2005}%
  \BibitemOpen
  \bibfield  {author} {\bibinfo {author} {\bibfnamefont {M.-P.}\ \bibnamefont {Valignat}}, \bibinfo {author} {\bibfnamefont {O.}~\bibnamefont {Theodoly}}, \bibinfo {author} {\bibfnamefont {J.~C.}\ \bibnamefont {Crocker}}, \bibinfo {author} {\bibfnamefont {W.~B.}\ \bibnamefont {Russel}},\ and\ \bibinfo {author} {\bibfnamefont {P.~M.}\ \bibnamefont {Chaikin}},\ }\bibfield  {title} {\bibinfo {title} {{Reversible self-assembly and directed assembly of DNA-linked micrometer-sized colloids}},\ }\href {https://doi.org/10.1073/pnas.0500507102} {\bibfield  {journal} {\bibinfo  {journal} {Proceedings of the National Academy of Sciences}\ }\textbf {\bibinfo {volume} {102}},\ \bibinfo {pages} {4225} (\bibinfo {year} {2005})}\BibitemShut {NoStop}%
\bibitem [{\citenamefont {Jacobs}\ and\ \citenamefont {Rogers}(2025)}]{Jacobs.2025}%
  \BibitemOpen
  \bibfield  {author} {\bibinfo {author} {\bibfnamefont {W.~M.}\ \bibnamefont {Jacobs}}\ and\ \bibinfo {author} {\bibfnamefont {W.~B.}\ \bibnamefont {Rogers}},\ }\bibfield  {title} {\bibinfo {title} {{Assembly of Complex Colloidal Systems Using DNA}},\ }\href {https://doi.org/10.1146/annurev-conmatphys-032922-113138} {\bibfield  {journal} {\bibinfo  {journal} {Annual Review of Condensed Matter Physics}\ }\textbf {\bibinfo {volume} {16}},\ \bibinfo {pages} {443} (\bibinfo {year} {2025})}\BibitemShut {NoStop}%
\bibitem [{\citenamefont {Sacanna}\ \emph {et~al.}(2010)\citenamefont {Sacanna}, \citenamefont {Irvine}, \citenamefont {Chaikin},\ and\ \citenamefont {Pine}}]{Sacanna.2010}%
  \BibitemOpen
  \bibfield  {author} {\bibinfo {author} {\bibfnamefont {S.}~\bibnamefont {Sacanna}}, \bibinfo {author} {\bibfnamefont {W.~T.~M.}\ \bibnamefont {Irvine}}, \bibinfo {author} {\bibfnamefont {P.~M.}\ \bibnamefont {Chaikin}},\ and\ \bibinfo {author} {\bibfnamefont {D.~J.}\ \bibnamefont {Pine}},\ }\bibfield  {title} {\bibinfo {title} {{Lock and key colloids}},\ }\href {https://doi.org/10.1038/nature08906} {\bibfield  {journal} {\bibinfo  {journal} {Nature}\ }\textbf {\bibinfo {volume} {464}},\ \bibinfo {pages} {575} (\bibinfo {year} {2010})}\BibitemShut {NoStop}%
\bibitem [{\citenamefont {Sacanna}\ \emph {et~al.}(2013)\citenamefont {Sacanna}, \citenamefont {Korpics}, \citenamefont {Rodriguez}, \citenamefont {Colón-Meléndez}, \citenamefont {Kim}, \citenamefont {Pine},\ and\ \citenamefont {Yi}}]{Sacanna.2013}%
  \BibitemOpen
  \bibfield  {author} {\bibinfo {author} {\bibfnamefont {S.}~\bibnamefont {Sacanna}}, \bibinfo {author} {\bibfnamefont {M.}~\bibnamefont {Korpics}}, \bibinfo {author} {\bibfnamefont {K.}~\bibnamefont {Rodriguez}}, \bibinfo {author} {\bibfnamefont {L.}~\bibnamefont {Colón-Meléndez}}, \bibinfo {author} {\bibfnamefont {S.-H.}\ \bibnamefont {Kim}}, \bibinfo {author} {\bibfnamefont {D.~J.}\ \bibnamefont {Pine}},\ and\ \bibinfo {author} {\bibfnamefont {G.-R.}\ \bibnamefont {Yi}},\ }\bibfield  {title} {\bibinfo {title} {{Shaping colloids for self-assembly}},\ }\href {https://doi.org/10.1038/ncomms2694} {\bibfield  {journal} {\bibinfo  {journal} {Nature Communications}\ }\textbf {\bibinfo {volume} {4}},\ \bibinfo {pages} {1688} (\bibinfo {year} {2013})}\BibitemShut {NoStop}%
\bibitem [{\citenamefont {Huang}\ \emph {et~al.}(2016)\citenamefont {Huang}, \citenamefont {Boyken},\ and\ \citenamefont {Baker}}]{Huang.2016}%
  \BibitemOpen
  \bibfield  {author} {\bibinfo {author} {\bibfnamefont {P.-S.}\ \bibnamefont {Huang}}, \bibinfo {author} {\bibfnamefont {S.~E.}\ \bibnamefont {Boyken}},\ and\ \bibinfo {author} {\bibfnamefont {D.}~\bibnamefont {Baker}},\ }\bibfield  {title} {\bibinfo {title} {{The coming of age of de novo protein design}},\ }\href {https://doi.org/10.1038/nature19946} {\bibfield  {journal} {\bibinfo  {journal} {Nature}\ }\textbf {\bibinfo {volume} {537}},\ \bibinfo {pages} {320} (\bibinfo {year} {2016})}\BibitemShut {NoStop}%
\bibitem [{\citenamefont {King}\ \emph {et~al.}(2012)\citenamefont {King}, \citenamefont {Sheffler}, \citenamefont {Sawaya}, \citenamefont {Vollmar}, \citenamefont {Sumida}, \citenamefont {André}, \citenamefont {Gonen}, \citenamefont {Yeates},\ and\ \citenamefont {Baker}}]{King.2012}%
  \BibitemOpen
  \bibfield  {author} {\bibinfo {author} {\bibfnamefont {N.~P.}\ \bibnamefont {King}}, \bibinfo {author} {\bibfnamefont {W.}~\bibnamefont {Sheffler}}, \bibinfo {author} {\bibfnamefont {M.~R.}\ \bibnamefont {Sawaya}}, \bibinfo {author} {\bibfnamefont {B.~S.}\ \bibnamefont {Vollmar}}, \bibinfo {author} {\bibfnamefont {J.~P.}\ \bibnamefont {Sumida}}, \bibinfo {author} {\bibfnamefont {I.}~\bibnamefont {André}}, \bibinfo {author} {\bibfnamefont {T.}~\bibnamefont {Gonen}}, \bibinfo {author} {\bibfnamefont {T.~O.}\ \bibnamefont {Yeates}},\ and\ \bibinfo {author} {\bibfnamefont {D.}~\bibnamefont {Baker}},\ }\bibfield  {title} {\bibinfo {title} {{Computational Design of Self-Assembling Protein Nanomaterials with Atomic Level Accuracy}},\ }\href {https://doi.org/10.1126/science.1219364} {\bibfield  {journal} {\bibinfo  {journal} {Science}\ }\textbf {\bibinfo {volume} {336}},\ \bibinfo {pages} {1171} (\bibinfo {year} {2012})}\BibitemShut {NoStop}%
\bibitem [{\citenamefont {Greef}\ and\ \citenamefont {Meijer}(2008)}]{Greef.2008}%
  \BibitemOpen
  \bibfield  {author} {\bibinfo {author} {\bibfnamefont {T.~F. A.~d.}\ \bibnamefont {Greef}}\ and\ \bibinfo {author} {\bibfnamefont {E.~W.}\ \bibnamefont {Meijer}},\ }\bibfield  {title} {\bibinfo {title} {{Supramolecular polymers}},\ }\href {https://doi.org/10.1038/453171a} {\bibfield  {journal} {\bibinfo  {journal} {Nature}\ }\textbf {\bibinfo {volume} {453}},\ \bibinfo {pages} {171} (\bibinfo {year} {2008})}\BibitemShut {NoStop}%
\bibitem [{\citenamefont {Aida}\ \emph {et~al.}(2012)\citenamefont {Aida}, \citenamefont {Meijer},\ and\ \citenamefont {Stupp}}]{Aida.2012}%
  \BibitemOpen
  \bibfield  {author} {\bibinfo {author} {\bibfnamefont {T.}~\bibnamefont {Aida}}, \bibinfo {author} {\bibfnamefont {E.~W.}\ \bibnamefont {Meijer}},\ and\ \bibinfo {author} {\bibfnamefont {S.~I.}\ \bibnamefont {Stupp}},\ }\bibfield  {title} {\bibinfo {title} {{Functional Supramolecular Polymers}},\ }\href {https://doi.org/10.1126/science.1205962} {\bibfield  {journal} {\bibinfo  {journal} {Science}\ }\textbf {\bibinfo {volume} {335}},\ \bibinfo {pages} {813} (\bibinfo {year} {2012})}\BibitemShut {NoStop}%
\bibitem [{\citenamefont {Eikelder}\ \emph {et~al.}(2019)\citenamefont {Eikelder}, \citenamefont {Adelizzi}, \citenamefont {Palmans},\ and\ \citenamefont {Markvoort}}]{Eikelder.2019}%
  \BibitemOpen
  \bibfield  {author} {\bibinfo {author} {\bibfnamefont {H.~M. M.~t.}\ \bibnamefont {Eikelder}}, \bibinfo {author} {\bibfnamefont {B.}~\bibnamefont {Adelizzi}}, \bibinfo {author} {\bibfnamefont {A.~R.~A.}\ \bibnamefont {Palmans}},\ and\ \bibinfo {author} {\bibfnamefont {A.~J.}\ \bibnamefont {Markvoort}},\ }\bibfield  {title} {\bibinfo {title} {{Equilibrium Model for Supramolecular Copolymerizations}},\ }\href {https://doi.org/10.1021/acs.jpcb.9b04373} {\bibfield  {journal} {\bibinfo  {journal} {The Journal of Physical Chemistry B}\ }\textbf {\bibinfo {volume} {123}},\ \bibinfo {pages} {6627} (\bibinfo {year} {2019})}\BibitemShut {NoStop}%
\bibitem [{\citenamefont {Klein}\ \emph {et~al.}(2018)\citenamefont {Klein}, \citenamefont {Perry},\ and\ \citenamefont {Manoharan}}]{Klein.2018}%
  \BibitemOpen
  \bibfield  {author} {\bibinfo {author} {\bibfnamefont {E.~D.}\ \bibnamefont {Klein}}, \bibinfo {author} {\bibfnamefont {R.~W.}\ \bibnamefont {Perry}},\ and\ \bibinfo {author} {\bibfnamefont {V.~N.}\ \bibnamefont {Manoharan}},\ }\bibfield  {title} {\bibinfo {title} {{Physical interpretation of the partition function for colloidal clusters}},\ }\href {https://doi.org/10.1103/physreve.98.032608} {\bibfield  {journal} {\bibinfo  {journal} {Physical Review E}\ }\textbf {\bibinfo {volume} {98}},\ \bibinfo {pages} {032608} (\bibinfo {year} {2018})},\ \Eprint {https://arxiv.org/abs/1806.00155} {1806.00155} \BibitemShut {NoStop}%
\bibitem [{\citenamefont {Curatolo}\ \emph {et~al.}(2023)\citenamefont {Curatolo}, \citenamefont {Kimchi}, \citenamefont {Goodrich}, \citenamefont {Krueger},\ and\ \citenamefont {Brenner}}]{curatolo.2023}%
  \BibitemOpen
  \bibfield  {author} {\bibinfo {author} {\bibfnamefont {A.~I.}\ \bibnamefont {Curatolo}}, \bibinfo {author} {\bibfnamefont {O.}~\bibnamefont {Kimchi}}, \bibinfo {author} {\bibfnamefont {C.~P.}\ \bibnamefont {Goodrich}}, \bibinfo {author} {\bibfnamefont {R.~K.}\ \bibnamefont {Krueger}},\ and\ \bibinfo {author} {\bibfnamefont {M.~P.}\ \bibnamefont {Brenner}},\ }\bibfield  {title} {\bibinfo {title} {{A computational toolbox for the assembly yield of complex and heterogeneous structures}},\ }\href {https://doi.org/10.1038/s41467-023-43168-4} {\bibfield  {journal} {\bibinfo  {journal} {Nature Communications}\ }\textbf {\bibinfo {volume} {14}},\ \bibinfo {pages} {8328} (\bibinfo {year} {2023})}\BibitemShut {NoStop}%
\bibitem [{\citenamefont {Holmes-Cerfon}\ \emph {et~al.}(2013)\citenamefont {Holmes-Cerfon}, \citenamefont {Gortler},\ and\ \citenamefont {Brenner}}]{Holmes-Cerfon.2013}%
  \BibitemOpen
  \bibfield  {author} {\bibinfo {author} {\bibfnamefont {M.}~\bibnamefont {Holmes-Cerfon}}, \bibinfo {author} {\bibfnamefont {S.~J.}\ \bibnamefont {Gortler}},\ and\ \bibinfo {author} {\bibfnamefont {M.~P.}\ \bibnamefont {Brenner}},\ }\bibfield  {title} {\bibinfo {title} {{A geometrical approach to computing free-energy landscapes from short-ranged potentials}},\ }\href {https://doi.org/10.1073/pnas.1211720110} {\bibfield  {journal} {\bibinfo  {journal} {Proceedings of the National Academy of Sciences}\ }\textbf {\bibinfo {volume} {110}},\ \bibinfo {pages} {E5} (\bibinfo {year} {2013})},\ \Eprint {https://arxiv.org/abs/1210.5451} {1210.5451} \BibitemShut {NoStop}%
\bibitem [{\citenamefont {Holmes-Cerfon}(2016)}]{Holmes-Cerfon.2016}%
  \BibitemOpen
  \bibfield  {author} {\bibinfo {author} {\bibfnamefont {M.}~\bibnamefont {Holmes-Cerfon}},\ }\bibfield  {title} {\bibinfo {title} {{Sticky-Sphere Clusters}},\ }\href {https://doi.org/10.1146/annurev-conmatphys-031016-025357} {\bibfield  {journal} {\bibinfo  {journal} {Annual Review of Condensed Matter Physics}\ }\textbf {\bibinfo {volume} {8}},\ \bibinfo {pages} {77} (\bibinfo {year} {2016})},\ \Eprint {https://arxiv.org/abs/1709.05138} {1709.05138} \BibitemShut {NoStop}%
\bibitem [{\citenamefont {Hübl}\ and\ \citenamefont {Goodrich}(2025)}]{Huebl.2025}%
  \BibitemOpen
  \bibfield  {author} {\bibinfo {author} {\bibfnamefont {M.~C.}\ \bibnamefont {Hübl}}\ and\ \bibinfo {author} {\bibfnamefont {C.~P.}\ \bibnamefont {Goodrich}},\ }\bibfield  {title} {\bibinfo {title} {{Accessing Semiaddressable Self-Assembly with Efficient Structure Enumeration}},\ }\href {https://doi.org/10.1103/physrevlett.134.058204} {\bibfield  {journal} {\bibinfo  {journal} {Physical Review Letters}\ }\textbf {\bibinfo {volume} {134}},\ \bibinfo {pages} {058204} (\bibinfo {year} {2025})}\BibitemShut {NoStop}%
\bibitem [{\citenamefont {Hübl}\ \emph {et~al.}(2025)\citenamefont {Hübl}, \citenamefont {Videbæk}, \citenamefont {Hayakawa}, \citenamefont {Rogers},\ and\ \citenamefont {Goodrich}}]{Huebl.2025b}%
  \BibitemOpen
  \bibfield  {author} {\bibinfo {author} {\bibfnamefont {M.~C.}\ \bibnamefont {Hübl}}, \bibinfo {author} {\bibfnamefont {T.~E.}\ \bibnamefont {Videbæk}}, \bibinfo {author} {\bibfnamefont {D.}~\bibnamefont {Hayakawa}}, \bibinfo {author} {\bibfnamefont {W.~B.}\ \bibnamefont {Rogers}},\ and\ \bibinfo {author} {\bibfnamefont {C.~P.}\ \bibnamefont {Goodrich}},\ }\bibfield  {title} {\bibinfo {title} {{The polyhedral structure underlying programmable self-assembly}},\ }\bibfield  {journal} {\bibinfo  {journal} {arXiv}\ }\href {https://doi.org/10.48550/arxiv.2501.16107} {10.48550/arxiv.2501.16107} (\bibinfo {year} {2025}),\ \Eprint {https://arxiv.org/abs/2501.16107} {2501.16107} \BibitemShut {NoStop}%
\bibitem [{Note1()}]{Note1}%
  \BibitemOpen
  \bibinfo {note} {Strictly speaking, the distribution is \protect \emph {geometric}, since $n$ is an integer and $1 \leq n < \infty $.}\BibitemShut {Stop}%
\bibitem [{\citenamefont {Brenner}(2017)}]{brenner2017}%
  \BibitemOpen
  \bibfield  {author} {\bibinfo {author} {\bibfnamefont {M.~P.}\ \bibnamefont {Brenner}},\ }\href {https://doi.org/10.1090/pcms/023} {\emph {\bibinfo {title} {{Mathematics and Materials: Ideas about Self-Assembly}}}},\ \bibinfo {series} {IAS/Park City Mathematics Series}, Vol.~\bibinfo {volume} {23}\ (\bibinfo  {publisher} {American Mathematical Society},\ \bibinfo {year} {2017})\BibitemShut {NoStop}%
\bibitem [{Note2()}]{Note2}%
  \BibitemOpen
  \bibinfo {note} {This remains true if the two-fold symmetry of a filament is exploited, which cuts down the required particle species by half, but leaves the prohibitive linear scaling unaffected.}\BibitemShut {Stop}%
\bibitem [{Note3()}]{Note3}%
  \BibitemOpen
  \bibinfo {note} {The length of $\langle n \rangle / (2m)$ is necessary, since there is a roughly 50\% chance a single-species segment binds with another one of the same species before encountering a filament of the subsequent species.}\BibitemShut {Stop}%
\bibitem [{\citenamefont {Hayes}\ \emph {et~al.}(2021)\citenamefont {Hayes}, \citenamefont {Partridge},\ and\ \citenamefont {Mirkin}}]{Hayes.2021}%
  \BibitemOpen
  \bibfield  {author} {\bibinfo {author} {\bibfnamefont {O.~G.}\ \bibnamefont {Hayes}}, \bibinfo {author} {\bibfnamefont {B.~E.}\ \bibnamefont {Partridge}},\ and\ \bibinfo {author} {\bibfnamefont {C.~A.}\ \bibnamefont {Mirkin}},\ }\bibfield  {title} {\bibinfo {title} {{Encoding hierarchical assembly pathways of proteins with DNA}},\ }\href {https://doi.org/10.1073/pnas.2106808118} {\bibfield  {journal} {\bibinfo  {journal} {Proceedings of the National Academy of Sciences}\ }\textbf {\bibinfo {volume} {118}},\ \bibinfo {pages} {e2106808118} (\bibinfo {year} {2021})}\BibitemShut {NoStop}%
\bibitem [{\citenamefont {Xiong}\ \emph {et~al.}(2009)\citenamefont {Xiong}, \citenamefont {Lelie},\ and\ \citenamefont {Gang}}]{Xiong.2009}%
  \BibitemOpen
  \bibfield  {author} {\bibinfo {author} {\bibfnamefont {H.}~\bibnamefont {Xiong}}, \bibinfo {author} {\bibfnamefont {D.~v.~d.}\ \bibnamefont {Lelie}},\ and\ \bibinfo {author} {\bibfnamefont {O.}~\bibnamefont {Gang}},\ }\bibfield  {title} {\bibinfo {title} {{Phase Behavior of Nanoparticles Assembled by DNA Linkers}},\ }\href {https://doi.org/10.1103/physrevlett.102.015504} {\bibfield  {journal} {\bibinfo  {journal} {Physical Review Letters}\ }\textbf {\bibinfo {volume} {102}},\ \bibinfo {pages} {015504} (\bibinfo {year} {2009})}\BibitemShut {NoStop}%
\bibitem [{\citenamefont {Lowensohn}\ \emph {et~al.}(2019)\citenamefont {Lowensohn}, \citenamefont {Oyarzún}, \citenamefont {Paliza}, \citenamefont {Mognetti},\ and\ \citenamefont {Rogers}}]{Lowensohn.2019}%
  \BibitemOpen
  \bibfield  {author} {\bibinfo {author} {\bibfnamefont {J.}~\bibnamefont {Lowensohn}}, \bibinfo {author} {\bibfnamefont {B.}~\bibnamefont {Oyarzún}}, \bibinfo {author} {\bibfnamefont {G.~N.}\ \bibnamefont {Paliza}}, \bibinfo {author} {\bibfnamefont {B.~M.}\ \bibnamefont {Mognetti}},\ and\ \bibinfo {author} {\bibfnamefont {W.~B.}\ \bibnamefont {Rogers}},\ }\bibfield  {title} {\bibinfo {title} {{Linker-Mediated Phase Behavior of DNA-Coated Colloids}},\ }\href {https://doi.org/10.1103/physrevx.9.041054} {\bibfield  {journal} {\bibinfo  {journal} {Physical Review X}\ }\textbf {\bibinfo {volume} {9}},\ \bibinfo {pages} {041054} (\bibinfo {year} {2019})},\ \Eprint {https://arxiv.org/abs/1902.08883} {1902.08883} \BibitemShut {NoStop}%
\bibitem [{\citenamefont {Gillespie}(2007)}]{Gillespie.2007}%
  \BibitemOpen
  \bibfield  {author} {\bibinfo {author} {\bibfnamefont {D.~T.}\ \bibnamefont {Gillespie}},\ }\bibfield  {title} {\bibinfo {title} {{Stochastic Simulation of Chemical Kinetics}},\ }\href {https://doi.org/10.1146/annurev.physchem.58.032806.104637} {\bibfield  {journal} {\bibinfo  {journal} {Annual Review of Physical Chemistry}\ }\textbf {\bibinfo {volume} {58}},\ \bibinfo {pages} {35} (\bibinfo {year} {2007})}\BibitemShut {NoStop}%
\bibitem [{\citenamefont {van Kampen}(1992)}]{Kampen1992}%
  \BibitemOpen
  \bibfield  {author} {\bibinfo {author} {\bibfnamefont {N.}~\bibnamefont {van Kampen}},\ }\href@noop {} {\emph {\bibinfo {title} {{S}tochastic {P}rocesses in {P}hysics and {C}hemistry}}}\ (\bibinfo  {publisher} {Elsevier Science Publishers, Amsterdam},\ \bibinfo {year} {1992})\BibitemShut {NoStop}%
\bibitem [{\citenamefont {Blatz}\ and\ \citenamefont {Tobolsky}(1945)}]{Blatz.1945}%
  \BibitemOpen
  \bibfield  {author} {\bibinfo {author} {\bibfnamefont {P.~J.}\ \bibnamefont {Blatz}}\ and\ \bibinfo {author} {\bibfnamefont {A.~V.}\ \bibnamefont {Tobolsky}},\ }\bibfield  {title} {\bibinfo {title} {{Note on the Kinetics of Systems Manifesting Simultaneous Polymerization-Depolymerization Phenomena}},\ }\href {https://doi.org/10.1021/j150440a004} {\bibfield  {journal} {\bibinfo  {journal} {The Journal of Physical Chemistry}\ }\textbf {\bibinfo {volume} {49}},\ \bibinfo {pages} {77} (\bibinfo {year} {1945})}\BibitemShut {NoStop}%
\bibitem [{\citenamefont {Stockmayer}(1943)}]{Stockmayer.1943}%
  \BibitemOpen
  \bibfield  {author} {\bibinfo {author} {\bibfnamefont {W.~H.}\ \bibnamefont {Stockmayer}},\ }\bibfield  {title} {\bibinfo {title} {{Theory of Molecular Size Distribution and Gel Formation in Branched-Chain Polymers}},\ }\href {https://doi.org/10.1063/1.1723803} {\bibfield  {journal} {\bibinfo  {journal} {The Journal of Chemical Physics}\ }\textbf {\bibinfo {volume} {11}},\ \bibinfo {pages} {45} (\bibinfo {year} {1943})}\BibitemShut {NoStop}%
\bibitem [{\citenamefont {Dongen}\ and\ \citenamefont {Ernst}(1984)}]{Dongen.1984}%
  \BibitemOpen
  \bibfield  {author} {\bibinfo {author} {\bibfnamefont {P.~G. J.~v.}\ \bibnamefont {Dongen}}\ and\ \bibinfo {author} {\bibfnamefont {M.~H.}\ \bibnamefont {Ernst}},\ }\bibfield  {title} {\bibinfo {title} {{Kinetics of reversible polymerization}},\ }\href {https://doi.org/10.1007/bf01011836} {\bibfield  {journal} {\bibinfo  {journal} {Journal of Statistical Physics}\ }\textbf {\bibinfo {volume} {37}},\ \bibinfo {pages} {301} (\bibinfo {year} {1984})}\BibitemShut {NoStop}%
\end{thebibliography}
\end{document}


\preprint{APS/123-QED}

\title{Supplementary Information: Stochastic size control of self-assembled filaments}

\author{Maximilian C. H\"ubl}
\author{Carl P. Goodrich}%
\affiliation{%
 Institute of Science and Technology Austria (ISTA), Am Campus 1, 3400 Klosterneuburg, Austria
}%

\maketitle
\section{Equilibrium calculations}
Here we provide a detailed derivation of the semiaddressable length distribution, Eq.~(2) of the main text, together with the complete expressions for the partition function, the first few moments of the distribution, and the relationship between the chemical potential and particle concentrations.

\subsection{Deriving the equilibrium length distribution}
As in the main text, we assume that all particle species are supplied at equal chemical potential $\mu$ and that all bonds between the same species have the same binding energy $\eself$, and that bonds between different species have binding energy $\eseries$.
Particles of species $\alpha$ may only bind to other particles of species $\alpha$, or to particles of species $\alpha + 1$.
Under these conditions, the partition function of any filament of length $n$ that consists of $k$ different single-species segments is given by
\begin{equation}\label{eq:Znk}
    Z(n, k) = V \Omega(n, k)\phi_0^n e^{\beta\left[n\mu + (n-k)\eself + (k - 1)\eseries\right]} \,,
\end{equation}
where $V$ is the system volume, $\phi_0$ is an arbitrary reference concentration, and $\Omega(n, k)$ is the rotational and vibrational partition function of the filament, which depends on the binding interactions between the particles.
If we neglect interactions between non-neighboring particles, then a filament is one-particle-reducible~\cite{Kardar.2007}, so that the entropy of a filament can be written as
\begin{equation}
    \Omega(n, k) = 8\pi^2 \, \omega^{n-1} \,,
\end{equation}
where $\omega$ is the per-bond entropy (assumed, for simplicity, to be the same for both bond types), and the factor of $8\pi^2$ is due to center of mass rotations of the entire chain (in three dimensions).
The per-bond entropy $\omega$, which depends on the microscopic interactions between the particles, can then simply be absorbed into the binding energies,
\begin{equation}
    \varepsilon_i \to \varepsilon_i + \frac{1}{\beta}\log(\omega \phi_0) \,,
\end{equation}
where $i=1,2$, making our description agnostic to the system-dependent microscopic details.

To obtain the partition function of all structures of length $n$, we need to count the ways in which a length-$n$ filament can be subdivided into $k$ segments, which is given by $\binom{n-1}{k-1}$.
Moreover, we need to keep in mind that the segments need not start at species 1 and end at species $m$, so we need to sum over all the $m - k + 1$ ways we can select $k$ contiguous species.
Combining these considerations with Eq.~\eqref{eq:Znk}, and defining $\de = \eseries - \eself$ and $\lambda = -\beta(\eself + \mu)$, gives the partition function for an arbitrary length-$n$ filament
\begin{equation}\label{eq:Zn}
    Z(n) = e^{-\lambda n} e^{-\beta\eseries} \sum_{k = 1}^m (m - k + 1) \binom{n-1}{k - 1} e^{k \beta\de} \,.
\end{equation}
In this expression and all other expressions below, we drop the explicit reference to the volume of the system $V$, the reference concentration $\phi_0$, and the factor $8\pi^2$ coming from rotations, as they cancel out of most relevant calculations.

The probability of observing a filament of length $n$ is proportional to $Z(n)$, so to obtain the length distribution, we need to compute the normalization constant
\begin{equation}
    Z = \sum_{n=1}^\infty Z(n) \,,
\end{equation}
which is the partition function of the whole system.

To this end, we first reorder the sums, so that we can perform the sum over $n$ first.
Looking at only the factors that depend on $n$ in Eq.~\eqref{eq:Zn}, we have
\begin{equation}
    z(k) = \sum_{n=k}^\infty \binom{n - 1}{k - 1} e^{-\lambda n} \,,
\end{equation}
where the lower summation bound starts at $k$, since a filament containing $k$ segments cannot have fewer than $k$ particles.
This sum can be performed with help from the generating function of the binomial coefficients, $\sum_{n=k}^\infty \binom{n}{k} e^{-\lambda n} = e^{-\lambda(k+1)} / (1 - e^{-\lambda})^k$.
Using this identity for our purposes, we obtain
\begin{equation}
    z(k) = \left[\frac{1}{e^{\lambda} - 1}\right]^k \,.
\end{equation}

Inserting this expression back into Eq.~\eqref{eq:Zn}, we have
\begin{equation}
    Z = e^{-\beta\eseries} \sum_{k=1}^m (m - k + 1) q^k \,,
\end{equation}
where we defined $q = e^{\beta\de}/(e^{\lambda} - 1)$ for convenience.
We can now easily perform the sum over $k$, and finally find
\begin{equation}
    Z = \frac{e^{\beta\mu}}{1 - e^{-\lambda}} \frac{q^{m+1} - (m+1)q + m}{(1 - q)^2} \,.
\end{equation}
Canceling the common factors contained in both $Z(n)$ and $Z$ from the above expression leads to the normalization constants $C_\mathrm{na}$ (if $m=1$) and $C_\mathrm{sa}$ (if $m>2$) of the main text.

\begin{widetext}
\subsection{Moments of the distribution}
With the expression for the partition function in hand, we can now compute the average length of the structures in the ensemble,
\begin{equation}
    \langle n \rangle = \frac{1}{Z} \sum_{n=1}^\infty Z(n) n \,.
\end{equation}
Using similar algebra as before, a somewhat lengthy calculation yields
\begin{equation}
    \langle n \rangle = \frac{mq^{m+2} + (m+2)(1 - q^m)q - m}{(q^{m+1} - (m+1)q + m)(1 - q)(1 - e^{-\lambda})} \,.
\end{equation}
Similarly, for the second moment,
\begin{equation}
    \langle n^2 \rangle = \frac{m(rq^{m+3} -rq^{m+2} + rq^2 - (r+1)q - 4q^{m+1} - 1)(q-1) - 2(rq^2 - (r+1)q - 2)(q^m - 1)q + m^2q^{m+1}(q-1)^2}{(q^{m+1} - (m+1)q + m)(1 - q)^2(1 - e^{-\lambda})^2} \,,
\end{equation}
where we also defined $r = e^{-\beta\de} / (1 - e^{-\lambda})$.
From this, we can compute the width of the distribution
\begin{equation}
    \sigma = \sqrt{\langle n^2 \rangle - \langle n \rangle^2} \,.
\end{equation}

\subsection{Computing particle concentrations}
As discussed above, we carry out our analytical calculations at uniform chemical potential for all particle species.
In order to compare our equilibrium calculations to the kinetic simulations, which are performed at fixed particle concentrations, we first need to relate the chemical potential $\mu$ to the concentrations.
To this end, we now compute the particle concentrations as functions of $\mu$.

The concentration $\rho_\alpha$ of particle species $\alpha \in (1, ..., m)$ is given by
\begin{equation}
    \frac{\rho_\alpha}{8\pi^2 \phi_0} = \sum_{s \in \mathcal{S}} n^s_\alpha Z(s) \,,
\end{equation}
where $s$ is a specific structure, $\mathcal{S}$ is the set of all structures, $n^s_\alpha$ is the number of particles of species $\alpha$ in $s$, and where we have restored the correct factors of $8\pi^2$ and $\phi_0$ that we dropped in the definition of $Z$.
Because $n^s_\alpha$ enters directly in this sum, we cannot immediately group structures by their total length, as we did above, and we need to be mindful of the structure compositions.

The trick to computing this sum is to split it into three parts: the section of the filament to the left of the species $\alpha$ whose concentration we want to compute, the section consisting only of species $\alpha$, and the section to the right of species $\alpha$.
We denote the length of the section to the left by $\ell_-$, and the number of species in that section by $k_-$.
Similarly, the length and number of species on the right are denoted by $\ell_+$ and $k_+$.
We also denote the number of particles of species $\alpha$ simply by $n$.
With this notation, the expression for $\rho_\alpha$ takes the form
\begin{equation}
     \frac{\rho_\alpha}{8\pi^2\phi_0} = \left[e^{-\beta \eself} \sum_{n=0}^\infty n e^{-\lambda n}\right] \times \left[\sum_{\ell_- = 0}^\infty e^{-\lambda\ell_-} \sum_{k_- = 0}^{\alpha - 1}\binom{\ell_- - 1}{k_- - 1} e^{\beta\de k_-}\right] \times \left[\sum_{\ell_+ = 0}^\infty e^{-\lambda\ell_+} \sum_{k_+ = 0}^{m - \alpha}\binom{\ell_+ - 1}{k_+ - 1} e^{\beta\de k_+}\right] \,,
\end{equation}
where independent sums have been separated by brackets.

The sums over $\ell_\pm$ and $k_\pm$ are almost identical to the sum we already computed above.
After a straightforward calculation, we find
\begin{equation}
     \frac{\rho_\alpha}{8\pi^2 \phi_0} = \frac{e^{\beta\mu}}{(1 - e^{-\lambda})^2} \times \frac{1 - q^\alpha}{1 - q} \times \frac{1 - q^{m - \alpha+1}}{1 - q} \,,
\end{equation}
where $q$ is defined as above and also depends on $\mu$.
Note that this expression is only valid for $1 < \alpha < m$.
If $\alpha = 1$, the second factor on the right-hand side needs to be replaced by $1$, as there cannot be a segment to the left of the first species.
Similarly, if $\alpha = m$, the last factor needs to be replaced by $1$.
This calculation shows that the concentrations of the different particle species are not necessarily uniform, even if the chemical potential is.
The particle numbers $N_\alpha = V \rho_\alpha$ and binding energies used in simulations are listed in Table~\ref{tab:sim_params}, showing quite strong variations between species.
For the simulations in Fig.~3 of the main text, we set the system volume to $V = 5 \times 10^8 / \phi_0$.

\begin{table}[]
    \centering
    \begin{tabular}{c|cccccccccccc}
        $\tilde{\sigma}$ & $N_1$ & $N_2$ & $N_3$ & $N_4$ & $N_5$ & $N_6$ & $N_7$ & $N_8$ & $N_9$ & $N_{10}$ & $\varepsilon_1 / \mathrm{kT}$ & $\Delta\varepsilon / \mathrm{kT}$ \\
        \hline
         1.0 & 4309579 & 5061737 & 5193009 & 5215899 & 5219776 & 5219776 & 5215899 & 5193009 & 5061737 & 4309579 & 29.62 & -0.50 \\
         1.25 & 3216626 & 4727217 & 5431933 & 5750711 & 5873512 & 5873512 & 5750711 & 5431933 & 4727217 & 3216626 & 21.74 & -1.65 \\
         1.50 & 2812744 & 4487013 & 5463659 & 5999841 & 6236743 & 6236743 & 5999841 & 5463659 & 4487013 & 2812744 & 20.20 & -2.04 \\
         1.75 & 2567754 & 4316742 & 5466439 & 6161116 & 6487949 & 6487949 & 6161116 & 5466439 & 4316742 & 2567754 & 19.44 & -2.34 \\
         2.0 & 2405066 & 4194799 & 5461839 & 6271773 & 6666523 & 6666523 & 6271773 & 5461839 & 4194799 & 2405066 & 18.97 & -2.61 \\
         2.25 & 2305717 & 4117180 & 5456651 & 6340591 & 6779861 & 6779861 & 6340591 & 5456651 & 4117180 & 2305717 & 18.66 & -2.90 \\
         2.50 & 2273016 & 4091140 & 5454565 & 6363435 & 6817845 & 6817845 & 6363435 & 5454565 & 4091140 & 2273016 & 18.45 & -3.20 
    \end{tabular}
    \caption{Parameters used in the kinetic simulations shown in Fig.~3 of the main text. $N_\alpha$ denotes the number of particles of species $\alpha$, $\eself$ is the same-species binding energy and $\Delta\varepsilon = \eseries - \eself$ is the difference between same-species and cross-species binding energies. Particle numbers and energies are chosen to reproduce a desired peak width $\sigma$, as described in the caption of Fig.3.}
    \label{tab:sim_params}
\end{table}
\newpage
\end{widetext}

\section{Assembly Kinetics}
\subsection{Kinetics of single-species filaments}
In order to set a baseline expectation we can compare against in the evaluation of equilibration times for multi-species assembly, we analytically compute the equilibration time of single-species polymerization, using the same assumptions we use for the general, multi-species case.
For this, we look at the deterministic version of the kinetic model we discuss in the main text with $m=1$ species, where the number densities $\rho_n$ of filaments of length $n$ obey the following infinite set of ordinary differential equations~\footnote{The notation in this section is somewhat different from the notation in the previous section in that the index $\rho_n$ now refers to the length of a filament, and not to the concentration of a specific particle species.}:
\begin{align}\label{eq:rates}
    \frac{d\rho_n}{dt} &= \frac{1}{2}\sum_{i=1}^{n-1} k(n-i, i) \rho_{n-i}\rho_i  - \rho_n\sum_{i = 1}^\infty k(n, i) \rho_i \nonumber \\
    &- \rho_n \sum_{i = 1}^{n - 1}f(n - i, i) + 2\sum_{i = 1}^\infty f(n, i) \rho_{n+i} \,.
\end{align}
Here, the first term corresponds to two shorter filaments combining to form a filament of length $n$, the second term corresponds to a length-$n$ filament combining with another, the third term corresponds to a length-$n$ filament breaking into two, and the last term corresponds to a filament breaking into two filaments, one of which has length $n$.
The kinetics is controlled by the aggregation rates $k(i, j)$ and fragmentation rates $f(i, j)$, which need to obey detailed balance conditions~\cite{Vigil.2009, Kampen1992}.
For simplicity, we assume that both fragmentation and aggregation rates are independent of filament length, and set $k(i, j) \equiv 2\alpha$ and $f(i, j) \equiv \delta$~\footnote{The notation in this section is different from the one in the previous section in that $\alpha$ does not refer to a specific particle species, but to the aggregation rate of filaments.}.

For this choice of rates, the rate equations are solvable analytically~\cite{Blatz.1945, Dongen.1984}.
However, since we are only interested in the equilibration time, a full solution is not required here.
To calculate the equilibration time, we simply sum Eq.~\eqref{eq:rates} over $n$, which will lead us to a closed equation for $\langle n \rangle(t)$.
We now perform this sum term by term.
\paragraph{First term.} We can visualize the summation by writing out the matrix $\rho_{n-i}\rho_i$, which gives
\begin{equation}
    \alpha \sum_{n=1}^\infty \sum_{i=1}^{n-1} \rho_{n-i} \rho_i = \alpha \sum_{i=1}^\infty \sum_{j=1}^\infty \left(\begin{array}{cccc}
          0 & 0 & 0 & \dots \\
         \rho_1^2 & 0 & 0 & \dots \\
         \rho_2 \rho_1 & \rho_1 \rho_2 & 0 & \dots\\
         \rho_3 \rho_1 & \rho_2^2 & \rho_1 \rho_3 & \dots \\
         \vdots & \vdots & \vdots & \ddots 
    \end{array}\right)_{ij} \,. \nonumber
\end{equation}
Clearly, the sum over column $j$ of this matrix is given by $\alpha \rho_j \sum_{i=1}^\infty \rho_i$.
The only problem is that each column is offset by one element from the previous.
In our case, however, this offset is irrelevant since mass conservation implies $\lim_{n\to\infty} \rho_n = 0$, which means that we can simply ignore the offset and have
\begin{equation}
    \alpha \sum_{n=1}^\infty \sum_{i=1}^{n-1} \rho_{n-i} \rho_i = \alpha \rho_\mathrm{s}^2 \,,
\end{equation}
where $\rho_\mathrm{s} = \sum_{i=1}^\infty \rho_i$ is the total number density of structures.
\paragraph{Second term.} The sum over this term can be performed immediately and gives
\begin{equation}
    - 2\alpha \sum_{n=1}^\infty \rho_n\sum_{i = 1}^\infty \rho_i = -2\alpha \rho_\mathrm{s}^2 \,. 
\end{equation}
\paragraph{Third term.} This sum can also easily be computed to give
\begin{equation}
    - \delta \sum_{n=1}^\infty (n-1) \rho_n =  -\delta(\rho_\mathrm{p} - \rho_\mathrm{s}) \,,
\end{equation}
where $\rho_\mathrm{p} = \sum_{n=1}^\infty n \rho_n$ is the total particle concentration.
\paragraph{Fourth term.} Finally, the fourth term is also readily summed by noting that each $\rho_n$ appears $n - 1$ times in the double sum:
\begin{equation}
    2\delta \sum_{n=1}^\infty\sum_{i = 1}^\infty \rho_{n+i} = 2\delta \sum_{n=1}^\infty (n - 1) \rho_n = 2\delta(\rho_\mathrm{p} - \rho_\mathrm{s}) \,.
\end{equation}

Since all four terms only depend on $\rho_\mathrm{s}$, which is related to the average length via $\langle n \rangle = \rho_\mathrm{p}/\rho_\mathrm{s}$, we can write down a closed differential equation for $\langle n \rangle$ alone:
\begin{equation}
    \frac{d\langle n \rangle}{dt} = \alpha \rho_\mathrm{p} - \delta(\langle n \rangle^2 - \langle n \rangle) \,.
\end{equation}
The solution to this equation with initial condition $\langle n \rangle(0) = 1$ is
\begin{equation}
    \langle n \rangle(t) = \frac{1}{2} + \frac{k}{\delta} \tanh\left[kt + \mathrm{atanh}\left(\frac{\delta}{2k}\right)\right] \,,
\end{equation}
with a rate constant
\begin{equation}
    k = \sqrt{\left(\alpha\rho_\mathrm{p} + \frac{\delta}{4}\right)\delta} \,.
\end{equation}
This rate constant is the inverse of the equilibration time
\begin{equation}
    \tau = \frac{1}{k} \approx \frac{1}{\sqrt{\alpha\rho_\mathrm{p}\delta}} \,,
\end{equation}
where the approximate equality is valid if $\delta \ll \alpha \rho_\mathrm{p}$\,, which corresponds to high binding energies and is an excellent approximation in our case.
The equilibrium length, which is reached as $t\to\infty$, is given by
\begin{equation}
    \langle n \rangle_\mathrm{eq} = \frac{1}{2} + \frac{k}{\delta} \approx \sqrt{\frac{\alpha\rho_\mathrm{p}}{\delta}} \,,
\end{equation}
where the approximate equality again assumes small $\delta$.
Using this expression for the equilibrium length, the equilibration time can also be written as
\begin{equation}
    \tau \approx \frac{\langle n \rangle_\mathrm{eq}}{\alpha\rho_\mathrm{p}} \,.
\end{equation}

\section{Supplementary data}
\subsection{Supplementary data for filament kinetics}
\begin{figure*}
    \centering
    \includegraphics[width=0.5\linewidth]{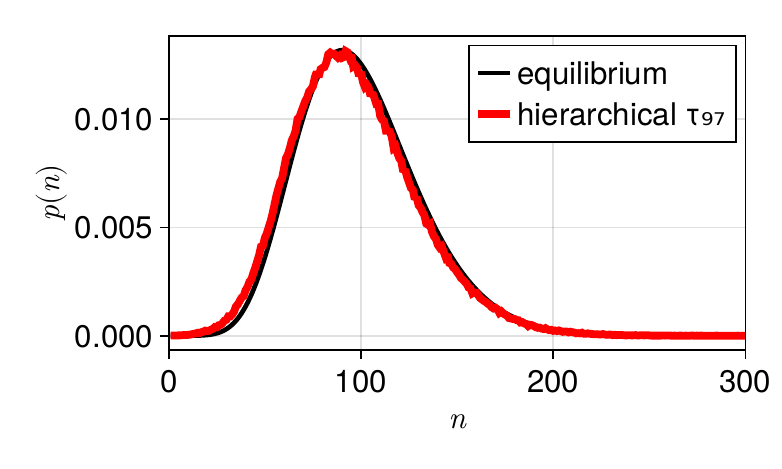}
    \caption{Length distribution during hierarchical assembly at time $t=\tau_{97}$ (red curve) and the corresponding equilibrium distribution (black curve).}
    \label{fig:figS1}
\end{figure*}

\paragraph*{Length distribution at $\tau_{97}$ --- }
Here we compare the filament length distributions for hierarchical assembly with the equilibrium distribution.
Figure.~\ref{fig:figS1} shows both length distributions obtained during hierarchical assembly at time $t = \tau_{97}$ and the equilibrium distribution, confirming that hierarchical assembly results in a distribution that is very close to equilibrium in a much shorter time compared to direct assembly.

\paragraph*{Stronger form of hierarchical assembly --- }
In addition to the hierarchical assembly protocol employed in the main text, we introduce here a stronger form of hierarchy, where filaments assemble at non-equilibrium conditions throughout.
In this strong hierarchy approach, we initially choose $\eself$ such that all $m$ species assemble single-species filaments of length $\langle n \rangle / m$.
After a time $t_\mathrm{h} \geq \tau_\mathrm{na}$, we freeze the momentary distribution of single-species filaments, meaning that existing filaments can neither break nor aggregate anymore.
At the same time, we enable unbreakable cross-species interactions, such that the single-species filaments join together in sequence.

\begin{figure*}
    \centering
    \includegraphics[width=0.5\linewidth]{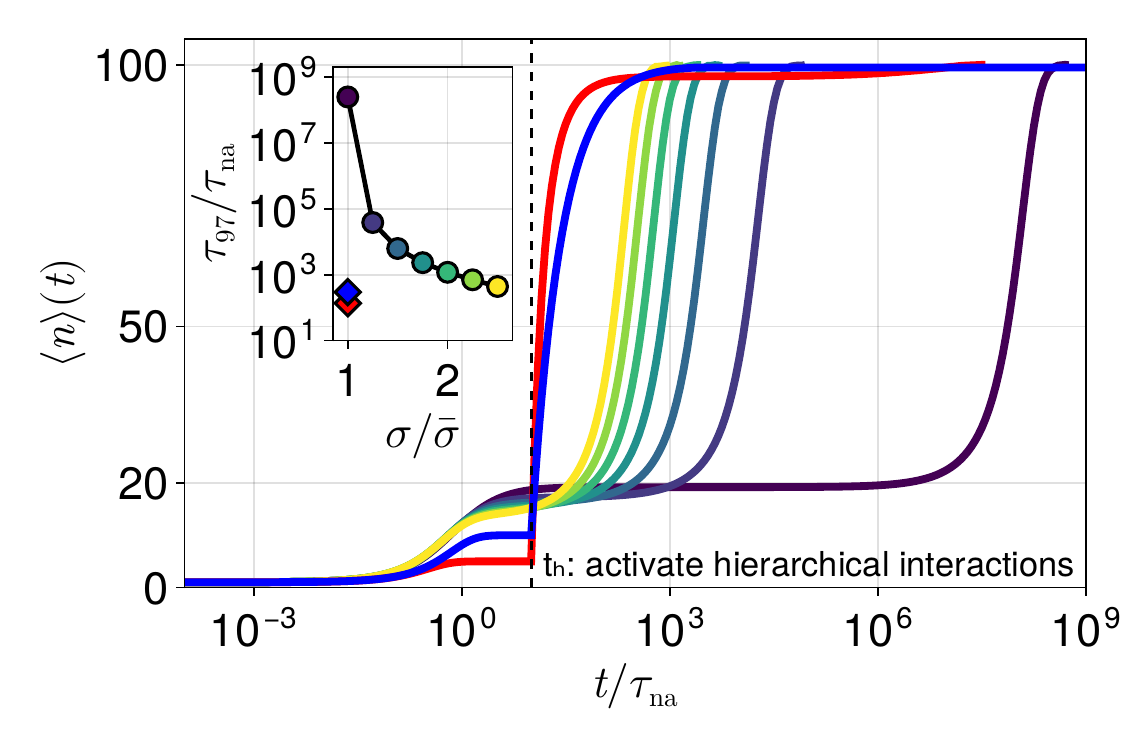}
    \caption{Kinetics of semiaddressable filaments with weak (red) and strong (blue) hierarchy. Time dependent average ensemble length for $m=10$ species and a target length of $\langle n \rangle(t = \infty) = 100$, for target peak widths $\sigma / \bar\sigma$ ranging from $1$ (dark purple) to $2.5$ (yellow), which are achieved by varying the binding energies. We use $\bar\sigma = \langle n \rangle / \sqrt{m}$, which is approximately equal to the peak width of the limiting distribution, as a reference. The red curve shows weak hierarchical assembly as discussed in the main text, the blue curve shows strong hierarchy.
    Both hierarchical protocols target $\sigma / \bar\sigma = 1$. Time is measured in units of the nonaddressable equilibration time $\tau_\mathrm{na}$.
    (\textit{inset}) Assembly time $\tau_{97}$ as a function of the target peak width for direct assembly (circles) and hierarchical assembly (diamond), showing the trade-off between assembly quality and time. For non-hierarchical systems, colors indicate peak width and are identical to the ones used in the main panel. Data for non-hierarchical and weak hierarchical assembly are identical to the data reported in Fig.~3 of the main text.}
    \label{fig:figS2}
\end{figure*}

Figure.~\ref{fig:figS2} shows a comparison of this strong hierarchy approach (blue curve) with the other data shown in Fig.~3 of the main text.
Interestingly, this strong version of hierarchy does not really perform better than the weaker, near-equilibrium version discussed in the main text: the assembly time $\tau_{97}$ for strong hierarchy is roughly two times higher compared to the weaker hierarchical protocol.
However, the stronger hierarchy does approach $\langle n \rangle = 100$ faster, so that any comparison between weak and strong hierarchy necessarily depends on the exact threshold at which the assembly is deemed good enough.

\subsection{Supplementary data for assembly of two-dimensional sheets}
As mentioned in \emph{Appendix}, we show in Fig.~\ref{fig:figS3} the size distribution of sheets assembled from a single filament species.
This distribution was obtained by simulating the assembly of $2\times 10^5$ filaments with fixed lengths drawn from a $m_1=10$ species distribution with target length $\langle n \rangle = 100$, and $\tilde\sigma = 1$, assembling at a vertical binding energy of $\varepsilon_{\mathrm{v}, 0} = 0.1 \,\mathrm{kT}$.
Interestingly, the dependence of the vertical binding energy on filament length causes the size distribution to develop multiple peaks, since longer filaments have a higher chance of forming ``taller'' sheets.

\begin{figure*}
    \centering
    \includegraphics[width=0.5\linewidth]{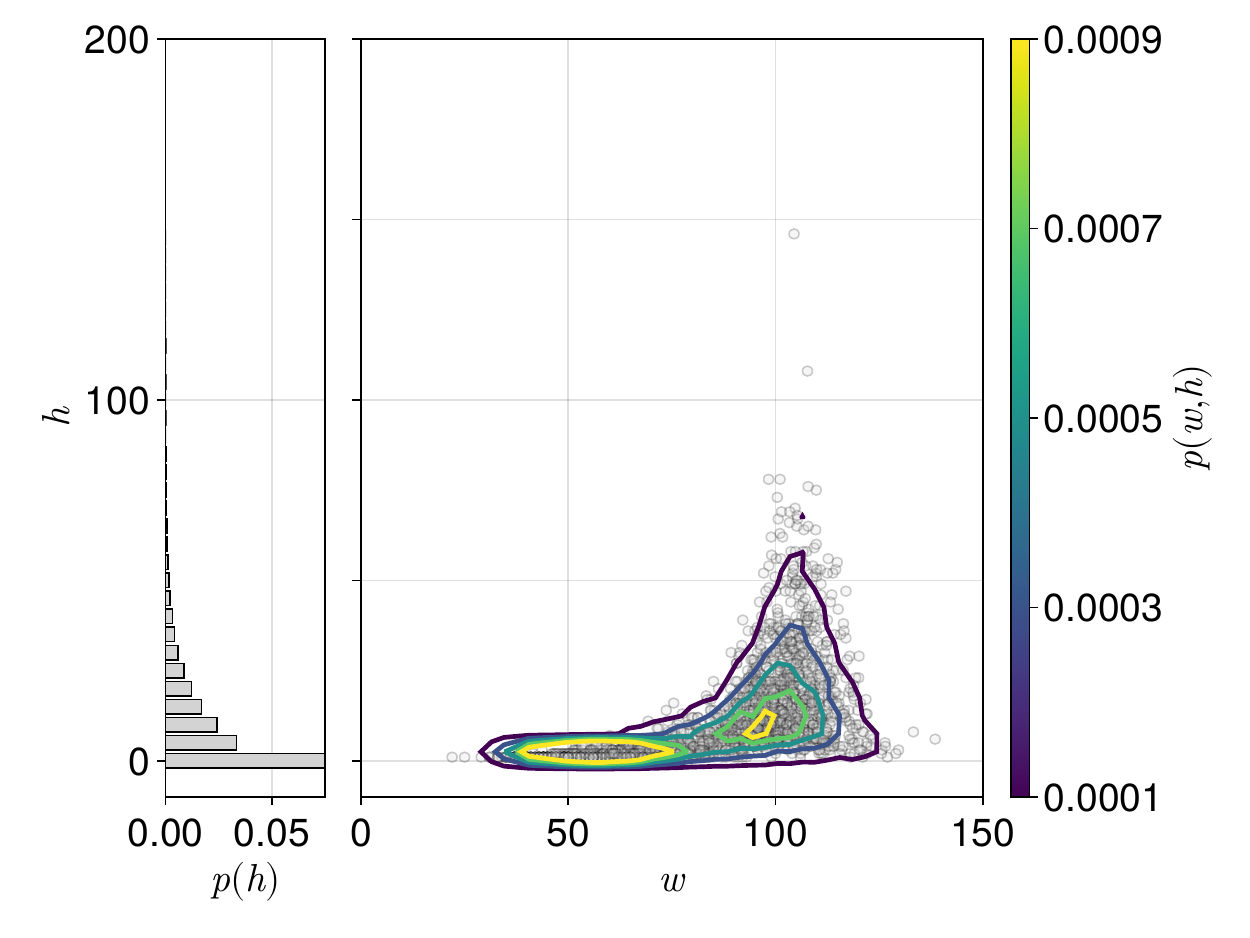}
    \caption{Size distribution of single-species two-dimensional sheets. The main panel shows the probability $p(w, h)$ of observing a sheet with width $w$ and height $h$, where width is defined as the mean length of the constituent filaments and height is given by the number of filaments in a sheet. Data points correspond to individual sheets and contour lines show a two-dimensional histogram. The panel on the left shows the marginal distribution of sheet heights $p(h)$.}
    \label{fig:figS3}
\end{figure*}